\documentclass[a4paper,10pt,oneside,onecolumn,nonatbib]{elsarticle}

\makeatletter
\def\ps@pprintTitle{%
 \let\@oddhead\@empty
 \let\@evenhead\@empty
 \def\@oddfoot{\centerline{\thepage}}%
 \let\@evenfoot\@oddfoot}
\makeatother
\usepackage{amsmath}
\usepackage[utf8]{inputenc}
\usepackage{cite}

\usepackage{tabularx}
\usepackage{booktabs}

\usepackage{siunitx}
\usepackage{comment}
\usepackage{graphicx}
\usepackage{dcolumn}
\usepackage{bm}
\usepackage[T1]{fontenc}
\usepackage{hyperref}
\usepackage[normalem]{ulem}
\usepackage{doi}
\usepackage{subcaption}
\usepackage{bbm}

\usepackage[acronym]{glossaries-extra}
\setabbreviationstyle[acronym]{long-short}
\glsdisablehyper

\usepackage[capitalise]{cleveref}

\newcommand\puteqnum{
  \refstepcounter{equation}\textup{(\theequation)}}

\newacronym{fee}{FEE}{front-end electronics}
\newacronym{daq}{DAQ}{data acquisition}
\newacronym{fov}{FOV}{field of view}
\newacronym{fcfov}{FCFOV}{fully coded field of view}
\newacronym[shortplural={PGs},longplural=prompt gammas]{pg}{PG}{prompt gamma}
\newacronym{pgh}{PG}{prompt-gamma} 
\newacronym{pgi}{PGI}{prompt-gamma imaging}
\newacronym{mlem}{MLEM}{maximum-likelihood expectation maximisation}
\newacronym[shortplural={SiPMs},longplural=silicon photomultipliers]{sipm}{SiPM}{silicon photomultiplier}
\newacronym{pet}{PET}{positron emission tomography}
\newacronym{mri}{MRI}{magnetic resonance imaging}
\newacronym{sificc}{SiFi-CC}{\textbf{Si}licon Photomultiplier and Scintillating \textbf{Fi}ber based \textbf{C}ompton \textbf{C}amera}
\newacronym{dfpd}{DFPD}{distal fall-off position determination}
\newacronym{dfp}{DFP}{distal fall-off position}
\newacronym{mura}{MURA}{modified uniformly redundant array}
\newacronym{cog}{CoG}{centre of gravity}
\newacronym{mps}{MPS}{multi-parallel slit}
\newacronym{kes}{KES}{knife-edge slit}
\newacronym[shortplural={ROIs},longplural=regions of interest]{roi}{ROI}{region of interest}
\newacronym{dpc}{DPC}{digital photon counter}
\newacronym{llr}{LLR}{low-level reconstruction}
\newacronym{snr}{SNR}{signal-to-noise ratio}
\newacronym{pmma}{PMMA}{poly(methyl methacrylate)}
\newacronym{ssp}{SSP}{small-scale prototype}
\newacronym{hit}{HIT}{Heidelberger Ionenstrahl-Therapiezentrum}
\newacronym{cm}{CM}{coded mask}
\newacronym{cmh}{CM}{coded-mask}
\newacronym{iqr}{IQR}{interquartile range}
\newacronym{rmse}{RMSE}{root mean squared error}
\newacronym{sm}{SM}{system matrix}
\newacronym{elar}{ELAR}{exponential light attenuation model with light reflection}

\newif\ifanonymous
\anonymousfalse

\begin{document}
\begin{frontmatter}

\title{First experimental test of a coded-mask gamma camera for proton therapy monitoring}

\ifanonymous

\else
\author[a2,a2a]{Magdalena~Ko\l{}odziej \texorpdfstring{\corref{cor1}}
}
\author[a4]{Stephan Brons}
\author[a2]{Miko\l{}aj Dubiel}
\author[a1]{George~N.~Farah}
\author[a1]{Alexander Fenger}
\author[a1]{Ronja~Hetzel\texorpdfstring{\fnmark[fn1]}{}}
\author[a1]{Jonas~Kasper}
\author[a2,a2a]{Monika~Kercz}
\author[a2,a2a]{Barbara~Ko\l{}odziej\texorpdfstring{\fnmark[fn2]}{}}
\author[a1]{Linn Mielke}
\author[a2]{Gabriel Ostrzo\l{}ek}
\author[a3]{Magdalena~Rafecas}
\author[a3]{Jorge~Roser}
\author[a2]{Katarzyna~Rusiecka \texorpdfstring{\corref{cor1}}}
\author[a1]{Achim~Stahl}
\author[a2]{Vitalii~Urbanevych}
\author[a2]{Ming-Liang~Wong\texorpdfstring{\fnmark[fn3]}{}}
\author[a2]{Aleksandra~Wro\'nska}

\cortext[cor1]{Corresponding authors. E-mail addresses: mkolodziej@doctoral.uj.edu.pl (Magdalena Kołodziej), katarzyna.rusiecka@uj.edu.pl (Katarzyna Rusiecka).}

\fntext[fn1]{Present address: Biophysics Department, GSI Helmholtzzentrum für Schwerionenforschung GmbH, Darmstadt, Germany.}

\fntext[fn2]{Present address: Medical University of Gdańsk, Gdańsk, Poland}

\fntext[fn3]{Present address: Department of Physics, University of Liverpool, Liverpool, UK.}

\address[a2]{Marian Smoluchowski Institute of Physics, Jagiellonian University, Krak\'ow, Poland}
\address[a2a]{Doctoral School of Exact and Natural Sciences, Jagiellonian University, Krak\'ow, Poland}
\address[a4]{Heidelberg Ion Therapy Center (HIT), Heidelberg, Germany}
\address[a1]{III. Physikalisches Institut B, RWTH Aachen University, Aachen, Germany}
\address[a3]{Institute of Medical Engineering, University of Lübeck, Lübeck, Germany}

\fi
\date{\today}

\begin{abstract}
\textbf{Objective.} 
The objective of the presented study was to evaluate the feasibility of a coded-mask (CM) gamma camera for real-time range verification in proton therapy, addressing the need for a precise and efficient method of treatment monitoring. \\
\textbf{Approach.}
A CM gamma camera prototype was tested in clinical conditions. The setup incorporated a scintillator-based detection system and a structured tungsten collimator. The experiment consisted of the irradiation of PMMA phantom with proton beams of energies ranging from 70.51 to \SI{108.15}{\mega\electronvolt}. Experimental data were benchmarked against Monte Carlo simulations. The distal falloff position was determined for both experimental data and simulations. \\
\textbf{Main results.}
The tested CM camera achieved a statistical precision of distal falloff position determination of \SI{1.7}{\milli\meter} for \num{e8} protons, which is consistent with simulation predictions, despite hardware limitations such as non-functional detector pixels. Simulations indicated that a fully operational setup would further improve the performance of the detector. The system demonstrated rate capability sufficient for clinical proton beam intensities and maintained performance without significant dead time. \\
\textbf{Significance.} 
This study validates the potential of the CM gamma camera for real-time proton therapy monitoring. The technology promises to enhance treatment accuracy and patient safety, offering a competitive alternative to existing approaches such as single-slit and multi-slit systems.
\end{abstract} 


\end{frontmatter}

\section{Introduction}

Proton therapy has become a well-established radiotherapy modality with more than 121 therapy centres currently operational and 30 under construction \cite{ptcog}. Irradiation with proton or heavy ion beams offers an advantageous dose deposition pattern and improved treatment conformity. However, large irradiation margins due to uncertainties in treatment planning, patient positioning, and physiological changes are still among the biggest problems for proton therapy, compromising treatment safety. The development of a method for online monitoring of proton therapy would allow for the reduction of the irradiation margins and thus improve the safety and quality of treatment. The need for such methods was one of the highlights of the report published by the Nuclear Physics European Collaboration Committee in 2014 \cite{NuPECC2014}.

Most approaches to real-time monitoring of proton therapy exploit by-products of patient irradiation, one of which is \gls{pgh} radiation. In 2006 it was experimentally proven that there is a correlation between the \gls{pgh} radiation characteristics and the position of the Bragg peak \cite{Min2006}. Since then, \gls{pgh} emission during proton therapy has been extensively studied \cite{Verburg2014, Pinto2015, Kelleter2017}. Several research groups have proposed detectors to determine the proton range or even the dose distribution based on the characteristics of \gls{pgh} radiation, such as timing, spatial distribution, and spectral characteristics. A recent review article by M. Pinto presents a comprehensive summary of the current status of \gls{pgh} based methods of proton therapy monitoring and challenges in the field \cite{Pinto2024}.

Many of the proposed detection setups incorporate passive collimation in various forms. The detection system featuring a \gls{kes} collimator developed by Richter et al. \cite{Richter2016} was the first one to be tested in clinical conditions and proven to provide control of inter-fractional range changes with a precision of about \SI{2}{\milli\meter}. In the second generation of this setup, the detector was mounted on a dedicated support frame, which was docked at a fixed position under the therapeutic table during patient irradiation. Systematic quantitative studies of the introduced improvements yielded an overall uncertainty of range prediction validation of about \SI{1}{\milli\meter} ($2\sigma$) \cite{Berthold2021}. The group recently reported that they achieved a reduction in irradiation margins from \SI{7}{\milli\meter} to \SI{3}{\milli\meter} in patients with prostate cancer using their second-generation \gls{kes} camera \cite{Bertschi2023}. The group of Xie et al. also tested a \gls{kes} detection system and proved its feasibility in clinical conditions in pencil beam scanning mode \cite{xie_prompt_2017}.

Another type of detection system which is considered for online range verification in proton therapy consists of the \gls{mps} collimator. Compared to \gls{kes}, it ensures a larger number of registered photons and offers larger \gls{fov}. \gls{mps} systems were tested by means of Monte Carlo simulations \cite{Pinto2014} as well as in experiments \cite{Park2019}. However, no superiority to the \gls{kes} systems was demonstrated at the time. The Lyon group has continued analytical modelling and Monte Carlo simulations of the two types of setups \cite{Huisman2023}. A recent article from Ku et al. reports on a \gls{mps} gamma camera that was tested under clinical conditions in spot-scanning mode. The group achieved precision ranging from \SI{2.2}{\milli\meter} to \SI{3.7}{\milli\meter} in range measurements for \num{e8} protons, depending on the beam energy. The group is planning to continue their studies with antropomorphic phantoms and patients \cite{Ku2023}.

Ready et al. proposed a variation of a \gls{mps}, featuring many knife-edge-shaped slits \cite{Ready2016, Ready2016phd}. The range retrieval precision achieved was \SI{1}{\milli\meter} ($2\sigma$); however, the tests were conducted with a beam energy of \SI{50}{\mega\electronvolt}, which is well below the clinically applicable energy range. Unfortunately, the studies of that group were discontinued. However, another team has adopted the concept and extended it to a dual-head configuration, enabling 3D imaging. The feasibility of using this setup for online range monitoring in proton therapy was demonstrated through simulations, with a position resolution better than \SI{2}{\milli\meter} across the whole \gls{fov}. The group is currently developing a prototype setup to verify their simulation results \cite{Lu2022}.

Yet another approach to passive collimation is a \gls{cmh} collimator. The \gls{cm} is an extension of the well-known pinhole camera concept. In contrast to a pinhole camera, in \gls{cmh} imaging the detector is shielded with a collimator consisting of many holes forming a specific pattern \cite{Cieslak2016}. When irradiated, the collimator casts a shadow on the detector surface. Depending on the position of the radiation source in \gls{fov}, the shadow is shifted. Knowing the distribution of hits recorded in the detector and the pattern of the mask, it is possible to reconstruct a 2D image of the radioactive source distribution. Using a \gls{cmh} collimator can be beneficial since even as many as half the pixels can be transparent to radiation, improving detection efficiency. At the same time, a mask pattern can be optimised such that the image resolution and \gls{snr} are also improved \cite{Fenimore1978, Gottesman1989}. A popular choice in \gls{cmh} imaging are \gls{mura} mask patterns \cite{Gottesman1989}. 

\gls{cmh} imaging was originally developed for astrophysics and was used to observe distant gamma-ray sources \cite{Ables1968, Dicke1968}. Nowadays, the application of \gls{cmh} imaging has broadened and includes localisation of radioactive materials e.g. in nuclear safety and security. With some adaptations of the \gls{cmh} detector, it is also possible to localise sources of neutrons of various energies, which is in particular of interest to those fields \cite{Cieslak2016}. However, it needs to be noted that all of the listed applications entail far-field imaging, rather than near-field, which is the subject of proton therapy monitoring. A setup featuring a pixelated detector and \gls{cm} with \gls{mura} pattern proposed for proton therapy monitoring was studied via Monte Carlo simulations by Sun et al. \cite{Sun2020}. The authors reported an accuracy of range determination better than \SI{0.8}{\milli\meter}. However, this result was obtained for \num{e10} impinging protons, which is two orders of magnitude more than is usually applied for a single distal irradiation spot. 

The study presented in this article is a continuation of the experiments previously conducted by our group  with small-scale prototypes of the \gls{cmh} camera and point-like radioactive sources \cite{Hetzel2023}. In that work, we proved that the \gls{cmh} technique can be successfully applied to the near-field imaging of gamma sources. We have also presented the results of Monte Carlo simulations of the full-scale setup featuring coded masks with a realistic source distribution resembling the conditions of proton therapy. At the statistics of \num{e8} impinging protons and in the beam energy range \SIrange{85.9}{107.9}{\mega\electronvolt}, we obtained the mean precision of beam range estimation of \SI{0.72}{\milli\meter} ($1\sigma$). 

The promising results of the pilot study led us to construct the full-scale \gls{cmh} detector. The experimental results obtained with the latter are the subject of this report. In the experimental test, the detector was used to detect range shifts in the phantom irradiated with proton beams of various energies. The experiments were carried out at the \gls{hit}. The coded mask pattern used and the overall performance of the sensitive detection part allowed for 1D imaging. In the following, we benchmark the obtained experimental results with Monte Carlo simulations. Furthermore, the simulation results show promising directions for further enhancement of the studied detector performance. 

The \gls{cm} camera described in the following article was developed under the \gls{sificc} project \cite{Kasper2020}. The goal of the \gls{sificc} collaboration is to construct a novel dual-modality setup for the online monitoring of proton therapy. One of the proposed detection modalities is a Compton camera \cite{Roser2024}, and the other is a \gls{cmh} camera. Both modalities are being developed in synergy, since they share most of their hardware, such as the sensitive material, \gls{fee}, and \gls{daq} system. After the necessary upgrade, the detector described in this article will serve in the future as a scatterer module of the full Compton camera proposed by the \gls{sificc} group. 

\section{Materials and methods}
\label{sec:materials-methods}

\subsection{Detection setup}
\label{sec:setup}
The investigated detection setup was of a \gls{cmh} camera type. The following sections detail its components. The selection of materials and the overall design were based on our earlier studies~\cite{Rusiecka2023, Hetzel2023}.

\begin{figure}
\centering
\includegraphics[width = 0.49\textwidth]{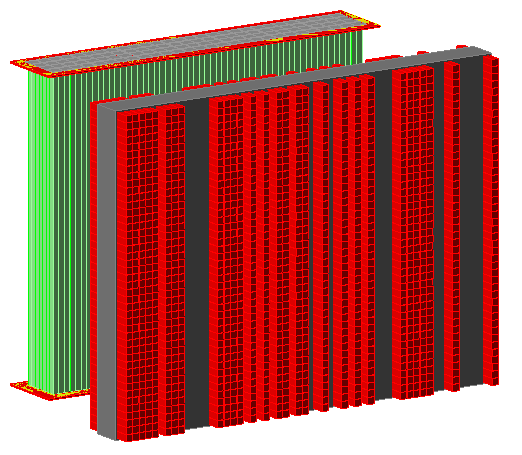}
\includegraphics[width = 0.49\textwidth]{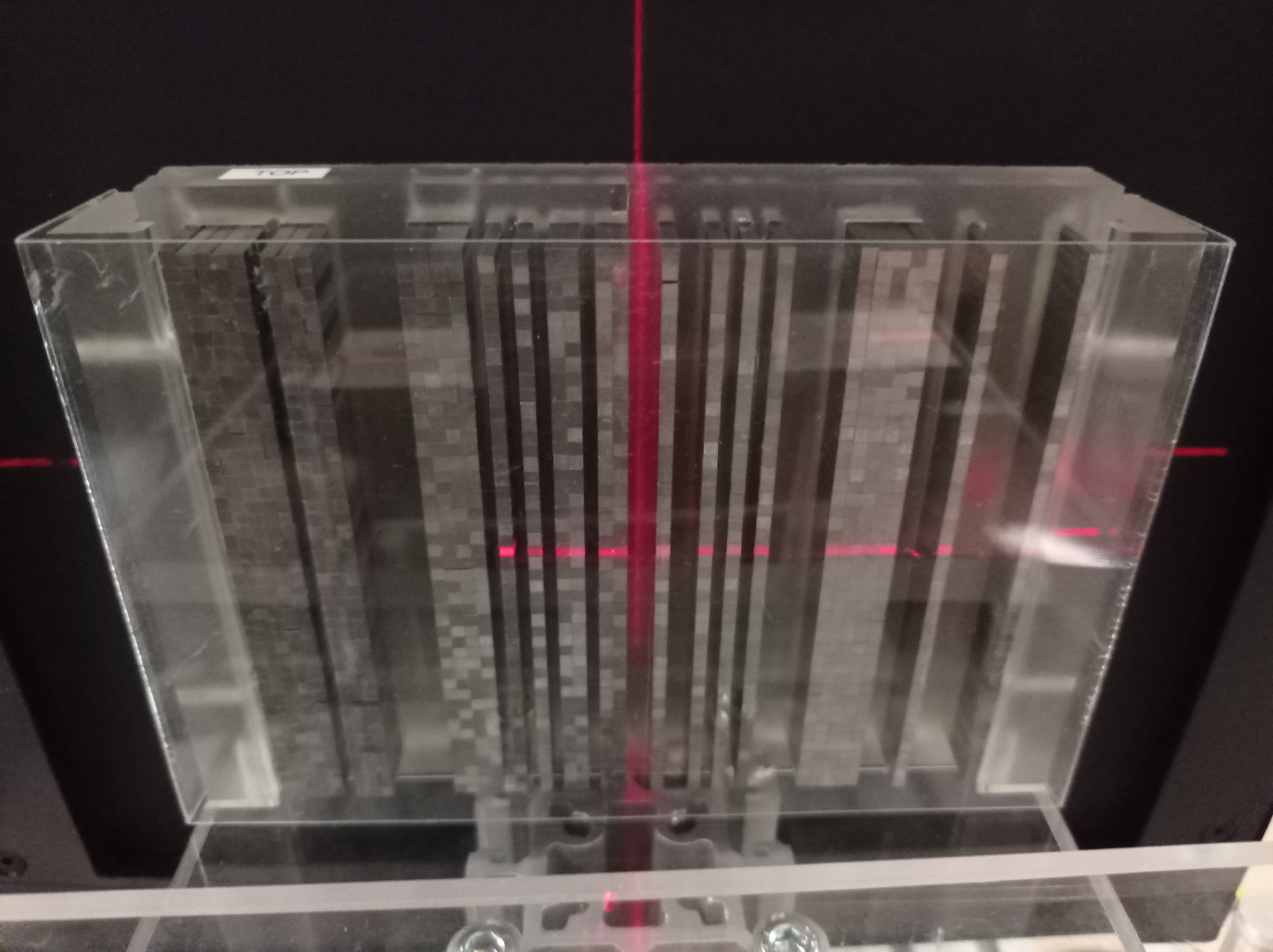}

\includegraphics[width = 0.9\textwidth]{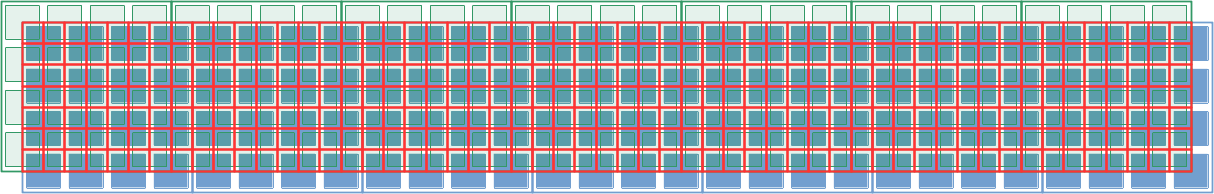}
\caption{Top left: a schematic of the detector setup: a stack of scintillation fibers (green) with dual top+bottom readout, and a structured collimator made of tungsten rods (red). Top right: collimator - real photo. Bottom: fiber and \gls{sipm} geometrical arrangement (top view). Red squares represent the fibers, blue and green boxes represent bottom and top \glspl{sipm}, respectively. }
\label{fig:CM1DSimPicture}
\end{figure}

\subsubsection{Sensitive part - scintillator}
\label{sec:scintillator}
The sensitive part was composed of elongated segments (termed \emph{fibers} in this work) of dimensions \qtyproduct[product-units=bracket-power]{1.94x1.94x100}{\mm}, made of  LYSO:Ce,Ca scintillation crystals. This material was chosen for its high effective atomic number and density, in order to enhance gamma detection efficiency. The fibers were arranged into an array of 7 layers, each layer comprising 55 fibers. Aluminium foil was used as optical insulation between the fibers to prevent optical crosstalk, resulting in an overall fiber pitch of the module of \SI{2}{\mm}. Both the scintillating material and the fiber wrapping material were chosen based on the extensive study~\cite{Rusiecka2021, Rusiecka2023}. The module was manufactured by Taiwan Applied Crystal~\cite{TAC}. 

\subsubsection{Front-end electronics and data acquisition system}
\label{sec:fee-daq}
The fiber stack was read out on two opposed sides by \glspl{sipm}, as depicted in~\cref{fig:CM1DSimPicture}. $4\times4$~arrays of the AFBR-S4N44P164M model from Broadcom~\cite{BroadcomArrays} were selected, as they had one of the highest photon detection efficiencies  (\SI{68}{\percent}) on the market at the time. Moreover, their spectral sensitivity matched well with the LYSO:Ce,Ca emission spectrum. The arrays were housed on custom printed circuit boards (PCBs) which also provided the hardware interface to the \gls{fee}. The \gls{sipm} form factor of the arrays was \SI{4}{\mm}, i.e. two times larger than that of the fibers in linear dimensions. This means that each \gls{sipm} collected light from four fibers. To facilitate identification of hit fibers in a way similar to that described in~\cite{LOPRESTI2014195}, the top and bottom \gls{sipm} boards were not mounted in a mirror configuration, but one was displaced diagonally with respect to the other by half of the \gls{sipm} pitch, as shown in the bottom panel of \cref{fig:CM1DSimPicture}. This reduced the number of required readout channels from 770 assuming the one-to-one fiber-\gls{sipm} coupling, to 224 with four-to-one coupling scheme.

Due to an incorrect method of soldering the \glspl{sipm} to PCB boards, some of them lost the electrical connection and were thus unusable during the experiment, causing acceptance gaps. Data analysis will exclude these non-working \glspl{sipm} and associated fibers. Henceforth, we refer to them as \emph{dead pixels}.

Fibers were coupled to \glspl{sipm} using a structured optical interface in the form of stainless steel, \SI{0.5}{\mm}-thick rasters with window patterns that matched those of the \gls{sipm} active areas. The function of the raster was to restrict the sensitivity of each \gls{sipm} to the fibers that directly faced it. The windows were filled with Elastosil RT 604 silicone rubber~\cite{Elastocil} to ensure good light transmission.

\gls{sipm} signals were further processed to extract the time and charge information in the TOFPET2c ASICs~\cite{DIFRANCESCO2016194} being parts of the integrated, off-the-shelf \gls{fee} and \gls{daq} system delivered by the PETSYS company~\cite{FEBD}. The system offers a possibility to trigger on events exhibiting a coincidence between two groups of channels, e.g. top and bottom \glspl{sipm} in that case. The used combination of PCIe with 3~SFP+ optical/copper connectors can transmit data up to \SI{6.6}{Gbit/s}. The selection of the \gls{fee}+\gls{daq} system was based on an extensive comparative study~\cite{Wong2024}. The coincidence time window of the \gls{daq} was set to \SI{15}{\ns} and the QDC integration time was \SI{290}{\ns}.

The masses of the detection module and the \gls{daq}~setup amounted to \SI{1.6}{\kilo\gram} and \SI{1.4}{\kilo\gram}, respectively. The light-tight box and the power supply are not included here.

\subsubsection{Mask}
\label{sec:mask}
Imaging experiments were performed using a collimator of the type of a \gls{mura} pattern. More specifically, we used a 476-rank 1D mask clipped to the central 57 pixels horizontally, and the pattern repeated over the 45 vertical pixel rows. This collimator is a slightly enlarged version of the one presented in the simulation study of~\cite{Hetzel2023} (51 horizontal pixels). The mask construction, as shown in the right panel of~\cref{fig:CM1DSimPicture} was based on a 3D-printed structured raster made of Phrozen Aqua Resin Clear, with pockets allowing to insert tungsten rods where filled pixels were foreseen. The rasters had a total thickness of \SI{13}{\mm} and the pockets to insert the rods were \SI{10}{\mm} deep. The mask pixel size was \SI{2.25}{\mm}, the precision of the rods manufacturing was \SI{0.05}{\mm}, and the thickness of the attenuating parts (i.e., the length of the rods) was \SI{20}{\mm}. The linear attenuation coefficient of tungsten amounts to \SI{0.784}{\per\cm}~\cite{nist}, i.e. \SI{79}{\percent} of \SI{4.4}{MeV} gammas are stopped in the mask thickness. A cover made of \SI{0.7}{\mm} thick \gls{pmma} preventing the rods from falling out was attached to the raster. The total mass of that structured collimator amounted to~\SI{3}{\kilo\gram}.

\subsection{Experiment}
\label{sec:experiment} 

\subsubsection{Conditions at HIT}
\label{sec:hit}
Measurements were performed in the experimental room of \gls{hit}. The facility offers various ion beam species, of which we exploited protons only. Importantly, the experimental room is equipped with a beam delivery system and a beam nozzle identical to those in the therapy rooms, which allows tests in truly clinical conditions. The machine offers predefined energy and intensity steps as well as several selectable lateral beam widths. Almost all measurements were carried out at the smallest beam width (ranging between \SI{14.5}{\mm} and \SI{22.5}{\mm} in the tested energy range) and the maximum available beam intensity step of 3.2$\times 10^9$~protons/s, although several measurements at lower beam intensity were taken for rate capability studies.

\subsubsection{Radioactive source}
\label{sec:source}
For the auxiliary measurements described in the following points, a $^{68}$Ge/$^{68}$Ga radioactive source was used. The source had a cylindrical shape of \SI{192}{\mm} length and \SI{3.2}{\mm} outer diameter, while the activity occupied the central part: \SI{184}{\mm} in length and \SI{1.6}{\mm} in diameter. The source decays via $\beta^{+}$ decay, but the shielding is penetrated only by the annihilation gamma quanta, with a small contribution of \SI{1077}{\keV} gammas. At the time of measurement, the source activity was \SI{10.7}{\mega\becquerel}.

\subsubsection{Energy- and \texorpdfstring{$y$}{y}-position calibration}
\label{sec:calib-measurements}
The light collection on the \glspl{sipm} depends on the position of the interaction along the fiber. This fact can be exploited to reconstruct the $y$-position based on the ratio of signals from both sides of the detector. However, this also implies that the energy calibration is $y$-dependent. Therefore, a series of calibration measurements was taken. In each of them, the source (described in \cref{sec:source}) was placed horizontally, i.e. along the $x$-axis, at different, known positions along the scintillating fibers. An additional reference detector, in the form of a horizontally oriented LYSO:Ce fiber with dual \gls{sipm} readout, was used in coincidence with the main detection module to constrain the direction of the gammas interacting with the latter, and thus their interaction position, taking advantage of the colinearity of a gamma pair emitted in a single act of annihilation. This principle is known as electronic collimation~\cite{Anfre2007, Rusiecka2023}. The AND signal of the two reference detector ends served as the \gls{daq} trigger. The vertical position of the source and reference detector was controlled remotely. The use of the electronic collimator instead of a passive one ensured better source positioning with the irradiated length of the detector defined by the source - reference detector distance and the source - module distance. The performed position scan consisted of nine 15-minute long measurements taken at \SI{10}{\milli\meter} intervals along the scintillating fibers, starting \SI{10}{\milli\meter} away from one end of the detector and finishing \SI{10}{\milli\meter} away from the opposite end.

\subsubsection{Experimental setup for imaging}
\label{sec:imaging-setup}
\begin{figure}[!ht]
\centering
\includegraphics[height = 5.3cm]{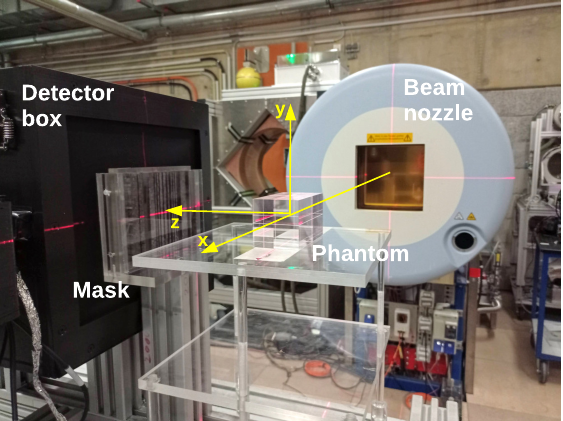}
\includegraphics[height = 5.3cm]{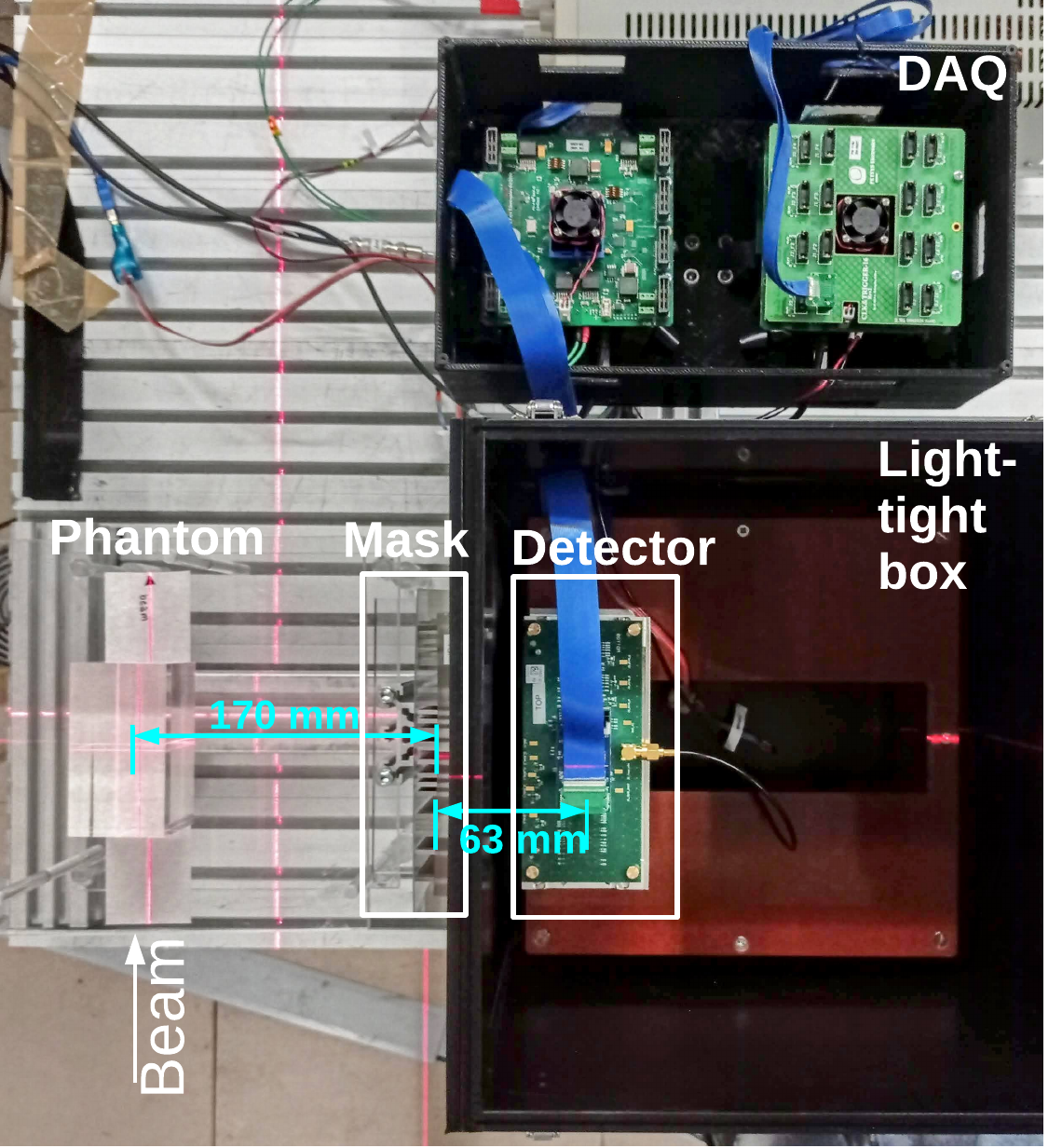}
\caption{Detection setup in imaging experiments: perspective view (left) and top view (right).}
\label{fig:SetupPhoto}
\end{figure}

The setup was arranged as shown in~\cref{fig:SetupPhoto}. A \gls{pmma} \qtyproduct[product-units=bracket-power]{50x50x90}{\mm} phantom with density of \SI{1.19}{\gram \per \cm^3}  was used as a target. The setup components were aligned with respect to the beam axis using the laser markers available in the experimental room, defining the beam axis and the isocentre. The longest axis of the phantom was aligned with the beam axis, the mask was located at a distance of \SI{170}{\mm} from it (distance to the mask centre), and the mask-detector distance was \SI{63}{mm} (centre-to-centre). Those distances are as close to the optimum found in~\cite{Hetzel2023} as was possible, given the constraints resulting from the presence of the light-tight box and components of the support structures. The setup geometry results in the \gls{fcfov} of \qtyproduct[product-units=bracket-power]{176x100}{\mm}, though in the following we restricted the analysis to \gls{fov} of \qtyproduct[product-units=bracket-power]{140x100}{\mm}.
As seen in \cref{fig:SetupPhoto}, the box housing the detector offered enough free space for convenient detector manipulation that takes place frequently in the prototyping phase. The \gls{daq}~electronics were located in a separate housing to reduce unwanted heat dissipation in the direct vicinity of the detector. 

\subsubsection{Proton beam measurements
\label{sec:HITmeas}}
\begin{figure}[!ht]
\centering
\includegraphics[width = 0.7\textwidth]{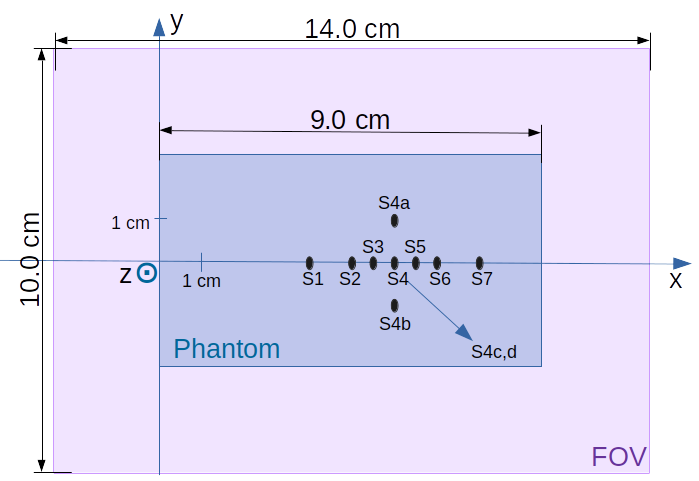}
\caption{Location of the irradiated spots in the phantom overlaid with the camera \gls{fov} - side view. Beam direction is $+\hat{x}$. For depth S4, data for four additional spots differing in lateral coordinates were taken. They are denoted with an additional character (e.g. S4c). Origin of the coordinate system, located in the phantom centre, is shown displaced for the sake of picture clarity.}
\label{fig:IrradiationPlan}
\end{figure}

We performed a series of measurements with different beam energies to test the beam range-shift retrieval capabilities of the setup. The corresponding beam energies, beam ranges as calculated using the PSTAR tool~\cite{pstar}, and beam lateral positions are presented in \cref{tab:ranges} along with the spot symbols referring to \cref{fig:IrradiationPlan}. The numbers of protons used for a single irradiation are also specified. The default number was $10^{10}$ protons, although for three spot locations ten times larger data samples were also recorded. Those measurements are referred to as primed (e.g. S3').

\begin{table}[ht]
  \centering
  \caption{Energies and ranges (calculated using PSTAR~\cite{pstar}) of the beam used in the experiment. Spot symbols refer to \cref{fig:IrradiationPlan}.}
  \label{tab:ranges}
  \begin{tabularx}{1\linewidth}{Xcccl}
    \toprule
    Spot symbol & Beam energy & Range & Lateral position & N protons \\ 
    & [MeV] & [mm] & $(z, y)$ [mm] &\\
    \midrule
    S1 & 70.51 & 36.63 & (0,0) & $10^{10}$\\[0.75ex]
    S2(S2') & 81.20 & 45.88 &  (0,0)& $10^{10}$ ($10^{11}$)\\[0.75ex]
    S3(S3') & 86.14 & 50.99 &  (0,0)& $10^{10}$ ($10^{11}$)\\[0.75ex]
    S4(S4') & 90.86 & 56.09 &  (0,0)& $10^{10}$ ($10^{11}$)\\
    S4a & & &  (0,10)& $10^{10}$\\ 
    S4b & & &  (0,-10)& \\
    S4c & & &  (-10,0)& \\
    S4d & & &  (10,0)& \\[0.75ex]
    S5 & 95.40 & 61.19 &  (0,0)& $10^{10}$\\[0.75ex]
    S6 & 99.78 & 66.27 &  (0,0)& $10^{10}$\\[0.75ex]
    S7 & 108.15 & 76.46 &  (0,0)& $10^{10}$\\[0.75ex]
    \bottomrule
  \end{tabularx}
\end{table}

\subsubsection{Efficiency and background studies}
\label{sec:effbg}
The registered spatial distribution of \gls{pgh} interaction points can be distorted in two ways: by a non-uniform detector efficiency, as well as a background contribution from intrinsic lutetium activity in LYSO:Ce,Ca. To quantify both effects, additional measurements were performed. LYSO:Ce,Ca background was investigated in a dedicated measurement, with both the collimator and the radioactive source removed. For the efficiency studies, a measurement with the linear radioactive source and in the absence of the collimator was recorded. The source was oriented parallel to the beam axis, at a distance of \SI{233}{\milli\meter} from the detector centre.
Recorded data were background-subtracted and compared with the corresponding ones from simulations (see \cref{sec:SimulatedReference} and \cref{sec:ResDetEff}) to calculate the detection efficiency maps. 
Both the efficiency and the background measurements lasted \SI{5}{\minute}.

\subsection{Simulations}
\label{sec:simulations} 
To simulate the reactions of the protons inside the target, we used the 10.4.2 version of the Geant4 package~\cite{Geant4} and the \mbox{QGSP\_BIC\_HP\_EMZ} physics list~\cite{PhysicsListGuide}. This combination was found to deliver results closest to experimental ones as regards \gls{pg} production~\cite{Wronska2021}. Additionally, we used the GODDeSS package, which is an extension to Geant4 that manages optical physics in scintillators~\cite{GODDeSS}, to build scintillation volumes, and to process optical photons to generate signals in the detector.

\subsubsection{Setup response to \texorpdfstring{\gls{pgh}}{PG} radiation}
\label{sec:sim-detector-response}
To assess the performance of the \gls{cm} setup, we performed Monte Carlo simulations corresponding to the experimental conditions. The simulated setup was identical to the experimental one (see \cref{sec:HITmeas}). The simulation process was divided into two separate steps:
\begin{enumerate}
    \item A proton beam interacts with the phantom. Secondary particles are produced and propagated until they leave the target. The corresponding secondary phase-space files are stored.
    \item Gamma quanta from phase-space files are used as input for the simulation of interactions of the particles with the detector. The simulation tracks the optical photons generated in the detector fibers until they are registered by \glspl{sipm}. Additionally, specific \gls{sipm} properties like wavelength-dependent photon detection efficiency, cross talk, dark counts, recharge timing are modeled and taken into account to mimic the measured detector signals. The output files contain information about the \glspl{pg} that triggered the detector, the data registered by \glspl{sipm}, as well as Monte Carlo truth about the hits in the fibers.
\end{enumerate}

\subsubsection{Reference for efficiency studies}
\label{sec:SimulatedReference}
One of the components needed to obtain an efficiency map is a simulation of the interaction probabilities in the detector. The simulated setup was identical to the one in the efficiency measurement described in \cref{sec:effbg}. The number of simulated gammas from phase space files was \num{4e8} and the polar angle of the emitted gammas was restricted to $\theta < 35^\circ$ to speed up the simulations. At the same time, this limit of the $\theta$ angle ensured that the gammas emitted from every part of the source which were within the detector acceptance were represented in the data sample.
 
\subsubsection{System matrix}
\label{sec:SystemMatrix}
To reconstruct \gls{pg} depth profiles based on the \gls{cm} camera data we used \gls{mlem} algorithm, which requires information about the probability of registering a particle emitted from a specific position in the source plane in a specific part of the detector. This information is represented by the so-called \gls{sm}. The procedure is described in more detail in \cref{imagereco}. The system matrix was determined using the simulations of the \gls{cm} setup response. Here, the detector response to point-like sources of \glspl{pg} is of interest. Thus, instead of using the first step of the simulation framework, we used the General Particle Source. The \gls{fov} of \SI{140}{\milli\meter} was divided into 100~bins along the $x$-axis and a single bin in $y$ centred in the middle of the detector height. For each individual simulation, a gamma source was placed at the centre of the corresponding bin, and \num{6.5e6} gamma quanta were simulated. Each time, the angle of emission was restricted to $\theta < 50^\circ$, which was chosen to be large enough to cover the entire mask from each source position. The energy spectrum of the simulated gamma particles was obtained from the spot S4 phase space file. 

\subsection{Analysis chain}
\label{sec:AnalysisChain} 
\begin{figure}[!htb]
\centering
\includegraphics[width = 0.7\textwidth]{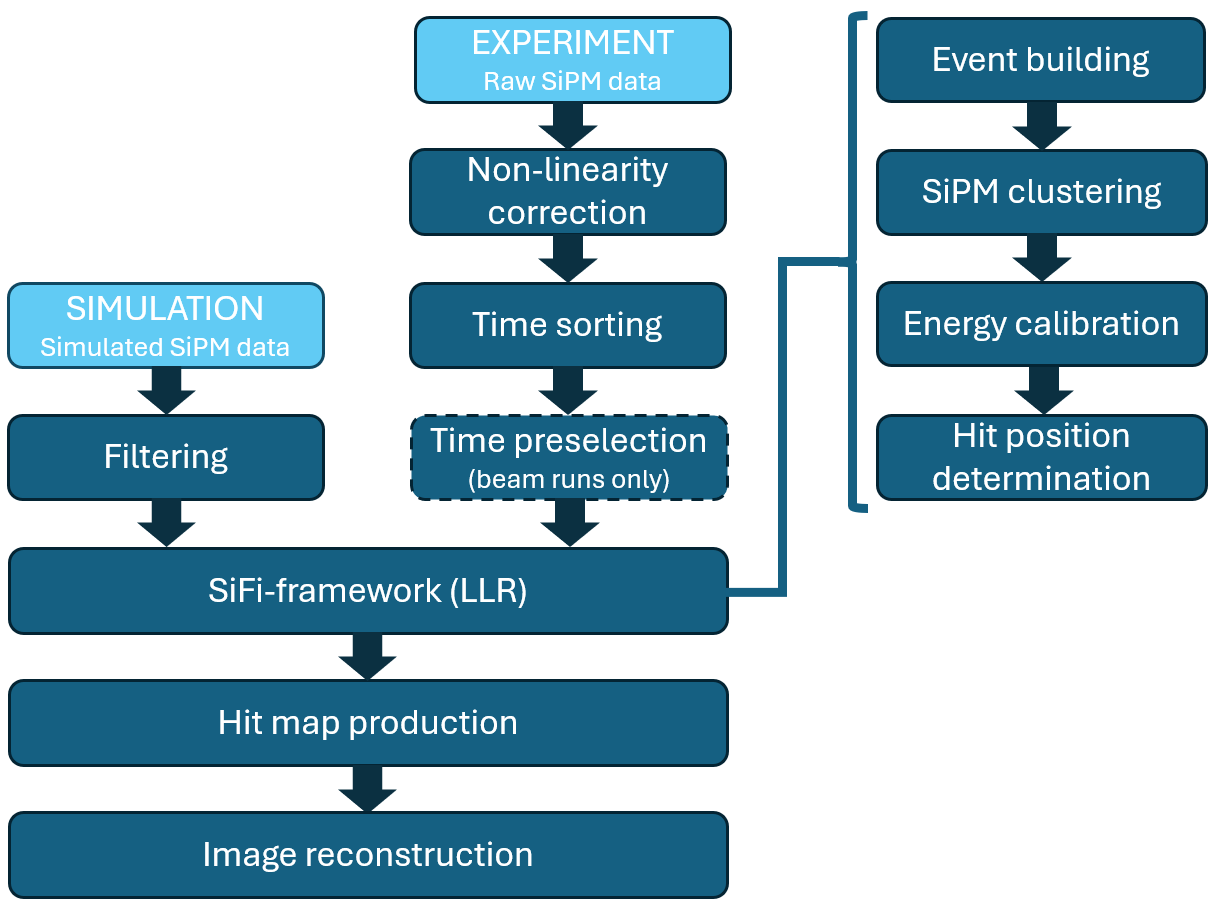}
\caption{The processing scheme of the experimental and simulated data.}
\label{fig:processingScheme}
\end{figure}

The complete data set for the evaluation of the performance of the \gls{cm} gamma camera contained: i) data for all beam spots in \cref{fig:IrradiationPlan}; ii) reference measurement without the \gls{cm} collimator with a linear $^{68}$Ge/$^{68}$Ga source for efficiency determination, and iii) background, measured at least 30 minutes after the last beam run, to eliminate the background component coming from activation of the setup parts by the beam. 
The data processing scheme is summarised in \cref{fig:processingScheme}.

The raw data collected in the experiment were saved in binary format. 
A single data entry was a set consisting of the ID of \gls{sipm} that registered a signal, the time of the signal registration, and the signal charge (QDC value) registered. Data were first corrected for a \gls{daq}-specific nonlinearity effect\cite{tofpetsoftwaremanual} and saved as ROOT trees \cite{root}. Then, they were sorted by time. 
Measurements with the beam were additionally divided into beam spill and inter-spill background parts. The trees processed in this way were then input into \gls{llr}, which consisted of event building, clustering, and reconstruction of hit positions and energy deposits in the detector. \gls{llr} was performed with custom software, the SiFi-framework~\cite{sifi-framework}.

We performed \gls{llr} on the simulation results in the same way we analysed the experimental data. In the case of the simulations, the data were already time-sorted and there was no need for \gls{daq}-specific non-linearity correction, but an extra step of filtering out inactive \glspl{sipm} was necessary. After that, the processing scheme was identical.

\subsubsection{Data preselection: signal and background subsets}
\label{preselection-signal-bg}
\begin{figure}[htb!]
\centering
\includegraphics[width = 0.65\textwidth]{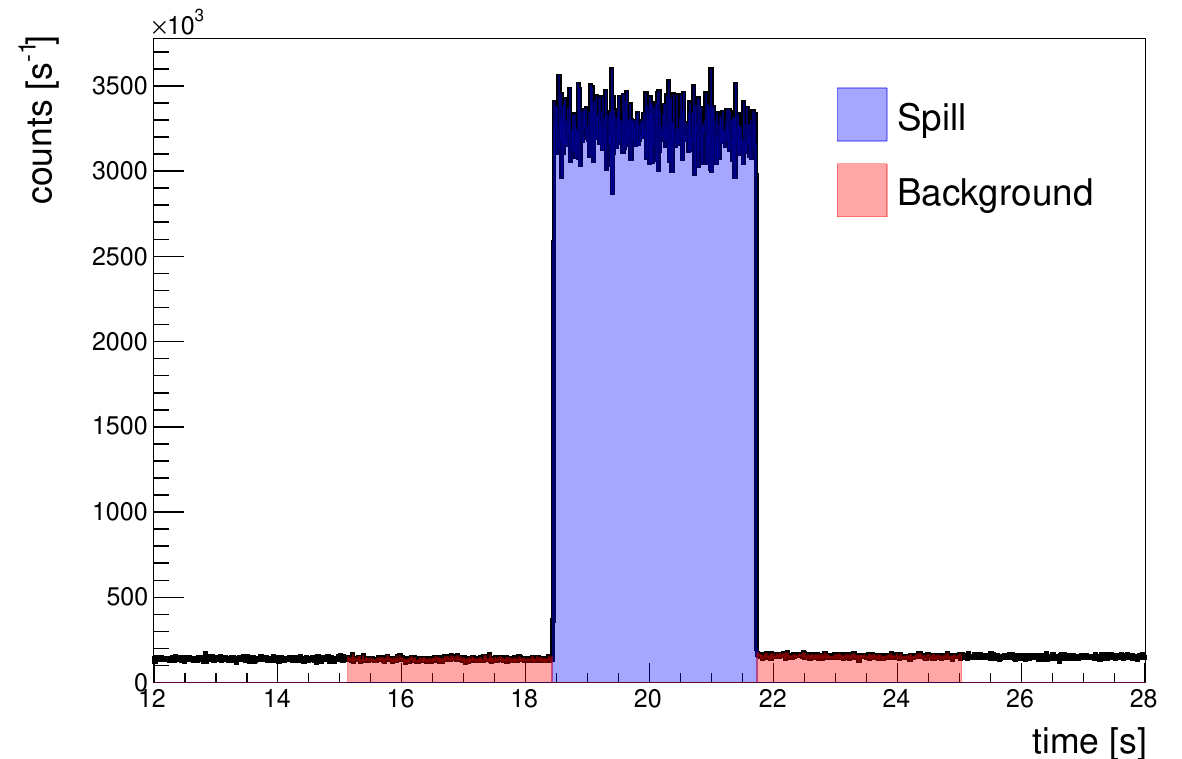}
\caption{Time structure of the proton beam, with separation to the spill (blue) and the background (pink) parts.}
\label{fig:TimeStructureOfTheBeam}
\end{figure}

\begin{figure}[htb!]
\centering
\includegraphics[width = 0.7\textwidth]{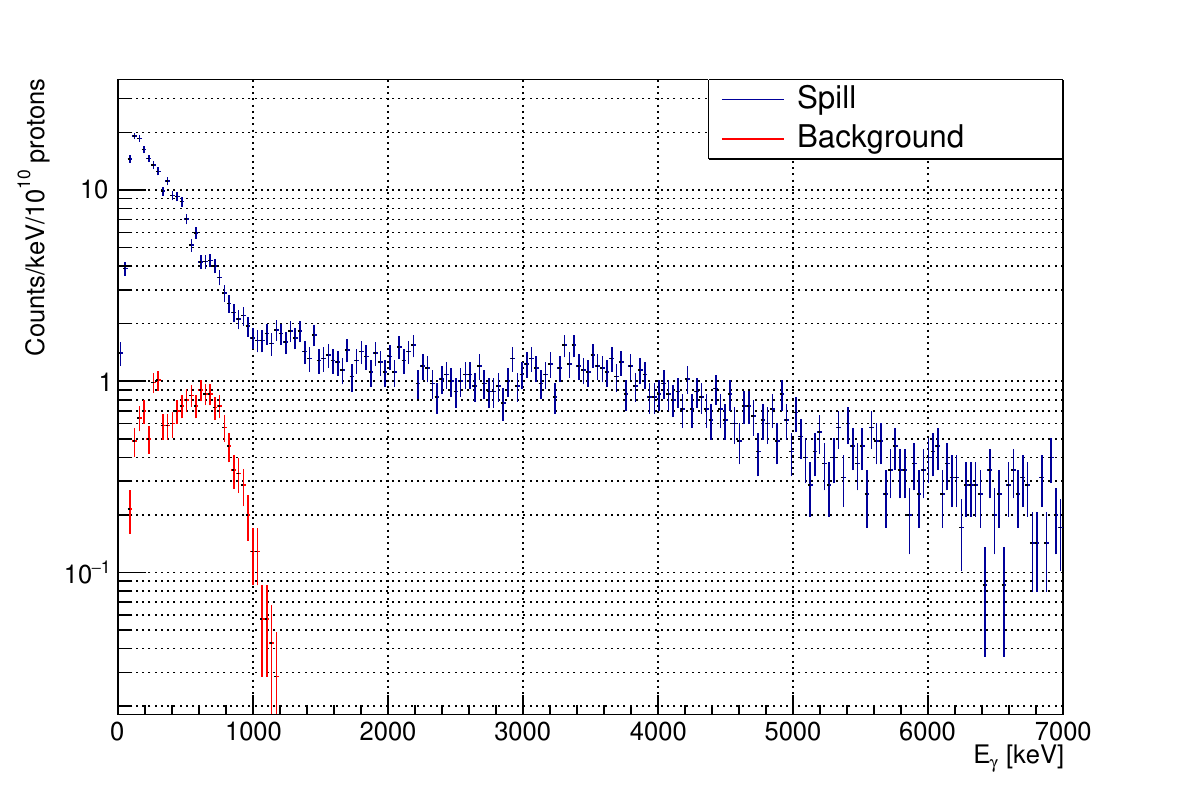}
\caption{Comparison of gamma energy spectra for spills and background for beam spot S4. The background was scaled to its expected content during the spill.}
\label{fig:beam_bg}
\end{figure}

The measurements with the beam needed an additional preprocessing step: time preselection. The event time distribution of an example run S4 presented in~\cref{fig:TimeStructureOfTheBeam} reflects the synchrotron beam time structure. One can see a region with increased counts corresponding to the beam spill, and an almost constant background. The data for each beam run were divided into spill and background subsets. The hit maps for these subsets are presented in~\cref{fig:allHitmaps} (a) and (b), respectively; their energy spectra are plotted in~
\cref{fig:beam_bg}. The background is considered during the image reconstruction procedure (see~\cref{imagereco}).

\subsubsection{Event building and clustering}
\label{sec:clustering}
The events were built out of single entries, based on time information: any entry that belonged to a fixed time window of \SI{15}{\nano\second} was assigned to the same event. The length of the window was inferred from a histogram of time difference between subsequent entries.

Within the events, \gls{sipm} clusters were found, separately for the top and bottom detector side. Then, for each pair of clusters, common fibers were found. We ensured that the fibers within those cluster pairs were grouped together, i.e., any fiber neighboured at least one other fiber. The events were classified as belonging to one of five categories, depending on the number of \gls{sipm} clusters; the abundance of the event types for run S4 (spill) is given in brackets:
\begin{enumerate}
    \item unique cluster, unique fiber: there is only one common fiber between the top and the bottom clusters (\SI{40.86}{\percent});
    \item unique cluster, multiple fiber: there is one top cluster and one bottom cluster, but more than one common fiber coupled to them (\SI{49.28}{\percent});
    \item top semi-unique cluster: two clusters on the bottom side, but one cluster on the top (\SI{5.58}{\percent});
    \item bottom semi-unique cluster: two clusters on the top side, but one cluster on the bottom (\SI{4.07}{\percent});
    \item ambiguous cluster: more than one cluster on both top and bottom sides (\SI{0.21}{\percent}).
\end{enumerate}

\subsubsection{Reconstruction of hit position and energy deposit}
\label{sec:hit-position-energy-reco}
The goal of \gls{llr} in our case was to obtain the hit positions in $x$ and $z$ dimensions for each event, along with the hit time and the energy deposit. 

The cluster hit time for all event types was the time of the first entry in the corresponding pair of \gls{sipm} clusters. 

Each of the event categories listed in \cref{sec:clustering} was handled differently as regards to the hit position and energy determination in \gls{llr}: 
\begin{itemize}
 \item The hit position for type-1 events was simply the position of the only fiber, and the energy deposit was the summed deposition in the top and bottom clusters.
 
 \item For type-2 events, we defined the hit position along the layer ($x$) as the mean of active fiber IDs, weighted by the fiber energy deposit (which was the geometric average of energy deposits in \glspl{sipm} coupled to a single fiber). The hit position value was rounded to the ID of the closest fiber. The hit position across the layers ($z$) was defined as the frontmost detector layer with an active fiber. Fiber positions were expressed by $f$ and $l$, i.e. the fiber number in the layer and the layer number, respectively. 
 The energy deposit for type-2 events was determined as $\sqrt{E_{T} E_{B}}$, where $E_{T}$, $E_{B}$ are summed energy deposits of all \glspl{sipm} forming the top ($T$) or bottom ($B$) cluster.  
 
 \item  Type-3 and type-4 events: prior to the hit position determination, one needs to separate the energy value of the merged cluster into two, according to the ratio of energies in clusters on the other side. The rest of the processing chain was the same as for type-2 events.
 
\item Type-5 events were excluded from further analysis.
\end{itemize}
Prior to the energy assignment, the QDC values for all \glspl{sipm} were calibrated with the use of an auxiliary measurement with a $^{68}$Ge/$^{68}$Ga radioactive source allowing one to find a scaling factor based on the \SI{511}{\keV} peak position. The scaling was then applied in \gls{llr}, to translate the QDC values to energy in keV.
 
 \subsubsection{Data representation: hit maps}
 \label{sec:hit-maps}
\begin{figure}[!htb]
\centering
\includegraphics[width =\textwidth]{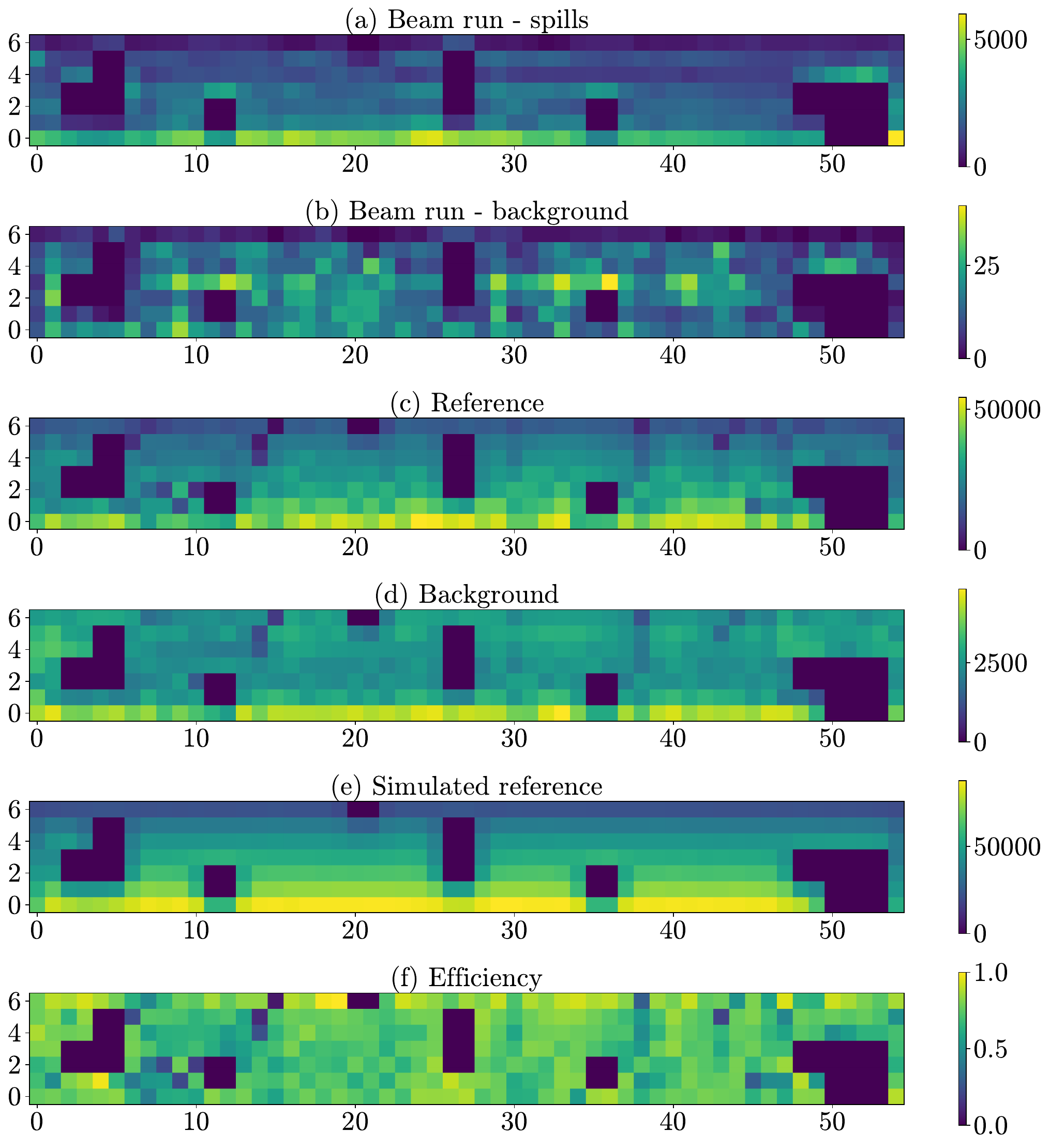}
\caption{Components of the image reconstruction, presented in the form of hit maps. On the hit map, the horizontal axis corresponds to the fiber number in the layer ($f$), with the  layer number ($l$) in the vertical axis.  a) spill part for S4; (b) background part for S4; (c)-(e) components of efficiency calculation: reference measurement, background and simulated reference, respectively; (f) efficiency calculated from components (c)-(e), as described in \cref{sec:ResDetEff}.}
\label{fig:allHitmaps}
\end{figure}

 As a next step, from the lists of fiber hits, we created maps of hits, as shown in \cref{fig:allHitmaps}. One pixel in such a map corresponds to one fiber, and each bin is filled with the number of hits registered in the corresponding fiber. The maps have dimensions of $7\times55$ pixels, which corresponds to spatial arrangement of fibers. We constructed hit maps with various lower energy thresholds (\num{0}, \num{0.5}, \num{1}, \SI{1.5}{\MeV}), and a common upper threshold of \SI{7}{\MeV}. Such hit maps were the input for image reconstruction, described in \cref{imagereco}.
 
\subsection{Image reconstruction}
\label{imagereco}
 
\subsubsection{Implementation of \texorpdfstring{\gls{mlem}}{MLEM} algorithm}
\label{sec:ImplementationOfMLEM}

To reconstruct the \gls{pg} depth profiles\footnote{In the following, we term the \gls{pg} depth profiles also \emph{images}, even though they are one-dimensional, for the method they are obtained (image reconstruction algorithm).} from the measured data, the \gls{mlem} \cite{Shepp1982, Lange1984} was implemented according to the following formula:
\begin{equation}
\label{eq:mlem}
    \bm{f}^{(k+1)} = \frac{\bm{f}^{(k)}}{\bm{S}} \textbf{A}^T \frac{ \bm{y}}{\textbf{A}\bm{f}^{(k)}+\bm{b}},
\end{equation}
where the vectors $\bm{f}^{(k)}$ and $\bm{f}^{(k+1)}$ represent the reconstructed \gls{pg} depth profiles  after iterations $k$ and $k+1$, respectively; the  element ${f}^{(k)}_j$ is thus the reconstructed number of emitted prompt gammas at the depth bin $j$, with $j \in \{1, 2\dotsc J\}$ and $J=100$;
the vector $\bm{y}$ corresponds to the histogrammed measured data (hit maps), while $\bm{b}$ is the estimated background; both vectors consist of 303 elements, i.e., the number of all detector pixels (385) minus the dead ones. The dead pixels were consistently removed from all other components of the MLEM formula.
The background  elements $b_i$ is a sum of what the detector pixel $i$ registers \SI{3}{\second} before and \SI{3}{\second} after the beam spill (see~\cref{fig:TimeStructureOfTheBeam}). The background was normalised to match the spill duration. 

\begin{figure}[!htb]
\centering
\includegraphics[width = 0.7\textwidth]{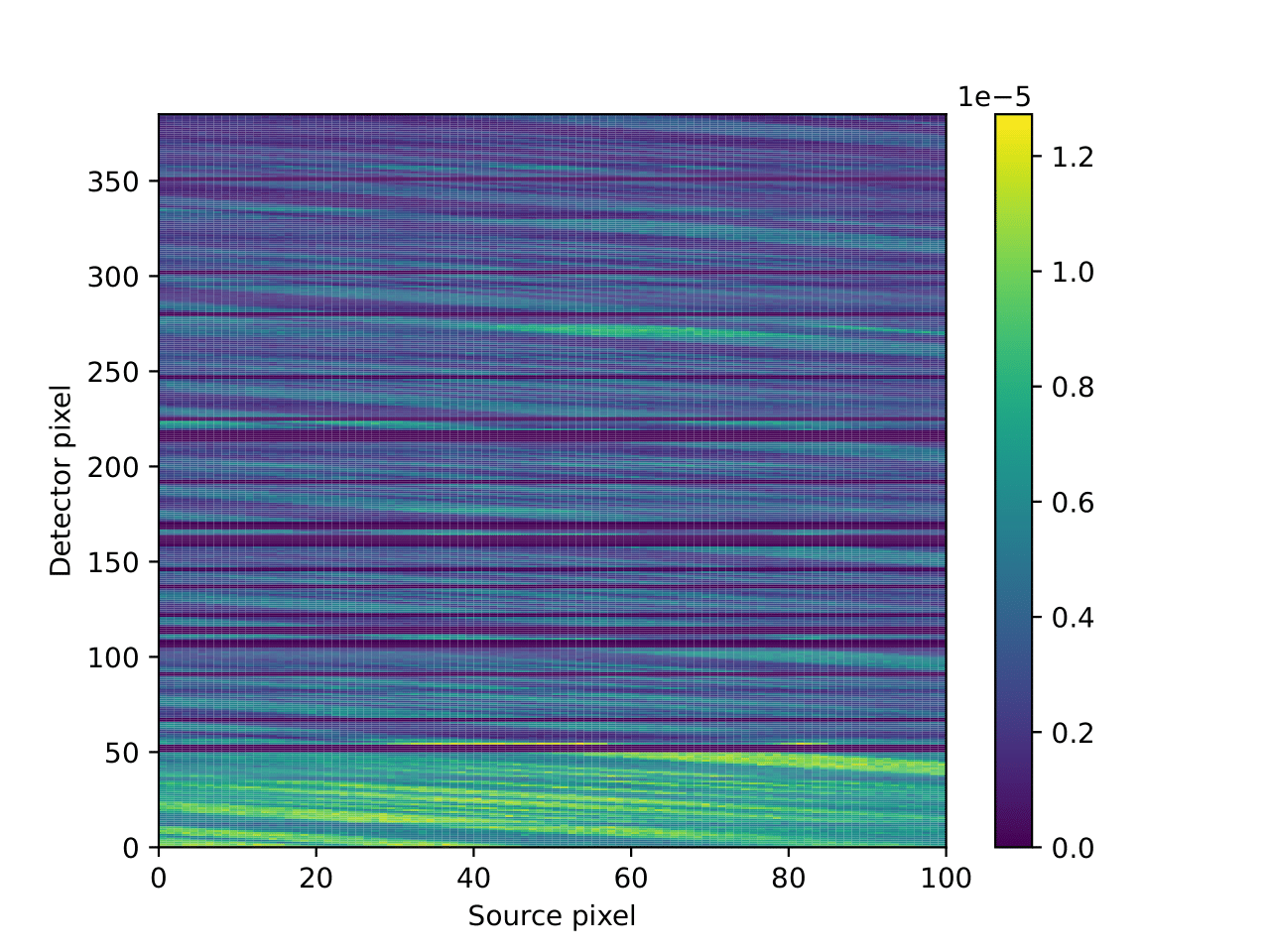}
\caption{The system matrix - constructed from hits with energy deposits \SIrange{1}{7}{\MeV}.
}
\label{fig:SM}
\end{figure}

$\textbf{A}$ denotes the system matrix, whose elements $a_{ij}$ represent the probability that a \gls{pg} emitted from the source bin $j$ is registered by the detector pixel $i$. 
First, we calculate $\tilde{a}_{ij}$, i.e., the ideal system matrix for uniform efficiency, using the simulation described in \cref{sec:SystemMatrix}: the recorded data are processed with the analysis chain described in~\cref{sec:AnalysisChain}), and 
for each source position $j$, the number of the registered gammas in detector pixel $i$ are computed. To convert this value into a probability, these values are divided by the number of gammas that would be emitted into the full solid angle in the performed simulation. 
The system matrix is presented in \cref{fig:SM}.

Differences in the detection efficiency between detectors pixels were also modeled into $\mathbf{A}$.
Efficiency variations are usually caused by hardware imperfections, such as flaws in the fibers, uneven or dirty coupling pads, machining of the \glspl{sipm} etc. To correct for these effects, two additional measurements of detector response to a linear $^{68}$Ge/$^{68}$Ga source ($R$) and to the pure LYSO:Ce,Ca background ($B$) were performed (see~\cref{sec:effbg}), along with the corresponding simulation of the detector response to the radioactive source ($SR$), see~\cref{sec:SimulatedReference}. 
For the detector element $i$, the efficiency correction factor was calculated as 
$\epsilon_i = (R_i- B_i)/{SR_i}$. Next,  the system matrix elements were calculated as $ a_{ij} = \epsilon_i \tilde{a}_{ij}$.

The elements of the sensitivity $\mathbf{S}$ were calculated as $S_j= \sum_{i=1}^{I} a_{ij}$.

\subsubsection{Determination of distal falloff position}
\label{sec:distal-falloff}
The crucial part of the \gls{pgh} depth profile, reconstructed according to the description in \cref{sec:ImplementationOfMLEM}, is the distal falloff. The \gls{dfp} is strongly correlated with the proton range in the tissue~\cite{Min2006}. We determined \gls{dfp} according to the procedure described in~\cite{Gueth2013}, by fitting a spline to the falling edge of the gamma profile. The fit range was set between the global maximum of the profile and a minimum further downstream the beam (in our representation: to the right of the maximum). As \gls{dfp}, we took the $x$ position of the point with half-value between the maximum and the minimum values. This value is subsequently compared with the proton range calculated using PSTAR~\cite{pstar} for \gls{pmma} and the proton beam energies used.

\subsubsection{MLEM performance metrics}
\label{sec:performanceMetrics}

To assess the image reconstruction performance, we defined three metrics based on the relation between the reconstructed \gls{dfp} and the calculated proton range PR: 
\begin{itemize}
    \item Pearson correlation coefficient;
    \item \gls{rmse}: \\
    \hfill 
      $\displaystyle     
      \hspace*{\fill}
      \mathrm{RMSE} = \sqrt{\frac{\sum_{i,j,i\neq j}\left( \Delta_{\mathrm{PR}}-\Delta_{\mathrm{DFP}}\right) ^2}{2n}}, \hfill \puteqnum \label{eq:oint} \\
      \hspace*{3.5cm} \Delta_{\mathrm{PR}} = \mathrm{PR}_{i}-\mathrm{PR}_{j},\\      
      \hspace*{3.5cm} \Delta_{\mathrm{DFP}} = \mathrm{DFP}_{i}-\mathrm{DFP}_{j},\\
      $
    where $n=21$ is the number of all pairs of measurements and $i,j\in [1,7]$ denote the number of beam spots. $\Delta$ denotes the difference between two distal falloff positions ($\Delta_{DFP}$) or two proton ranges ($\Delta_{PR}$). The additional factor of 2 in the denominator reflects the fact that RMSE is calculated based on two experimental points;
    \item slope of the linear fit to \glspl{dfp} versus proton range for all spots. 
\end{itemize}
 
In the optimisation process, we aimed for the correlation coefficient to be as high as possible, the RMSE as low as possible, and the slope preferably as close to one as possible. 

\subsubsection{Optimisation of image reconstruction parameters}
\label{sec:ParametersOptimization}

Using the metrics described above, the following reconstruction parameters were optimised:
\begin{itemize}
    \item number of iterations,
    \item threshold of energy deposit in hit maps,
    \item \gls{pg} depth profile smoothing method,
    \item excluded detector regions,
    \item classes of events included.
\end{itemize}

Using the performance metrics, the reconstruction was optimised and the best parameters were found.

\subsubsection{Analysis of statistical precision}
\label{sec:analysisOfStatisticalPrecision}

Data sets for all beam spots were randomly rearranged into 100 smaller subsets with the bootstrapping method. The sizes of the subsets corresponded to \num{2e9}, \num{1e9}, \num{4e8}, \num{2e8} and \num{1e8} protons on target. Then, the reconstruction with optimal parameters was performed on the subsets and, based on the distribution of the reconstructed DFPs, the statistical precision of the \gls{dfp} determination was assessed. 

\section{Results}

\subsection{Calibration results\label{sec:calres}}

The gamma energy spectra obtained in the calibration measurements for each fiber in the detector, featuring an annihilation peak, were later used for the calibration procedure. The calibration procedure developed by us in the single fiber tests~\cite{Rusiecka2021} and the tests with detector prototypes~\cite{Rusiecka2023} was applied. As a first step, the parameters of the \gls{elar} were fitted to the experimental data~\cite{Rusiecka2021}. Then, having QDC values recorded at both fiber ends, the interaction point coordinate along the fibers $(y)$ and the deposited energy can be reconstructed event by event. The obtained distributions yield the position- and energy resolution of the detector. In the presented experiment, the mean energy resolution was \SI{6.5(5)}{\percent} ($1\sigma$) and the mean position resolution was \SI{74(10)}{\milli\meter} (FWHM). The resolutions were determined based on the results obtained for about \SI{70}{\percent} of all fibers (excluding dead pixels and those for which the initial \gls{elar} model fit failed).

\subsection{Detection efficiency}
\label{sec:ResDetEff}

Determination of efficiency, as described in \cref{imagereco}, can be followed by analysing panels (c)-(f) of \cref{fig:allHitmaps}. The distortions in the direct neighbourhood of the dead pixel areas that are the artefacts of the \gls{llr} can be clearly seen in the efficiency map (f). After excluding the dead pixels, the relative standard deviation of fiber efficiency yielded 11.39\%. The mean efficiency (normalised to the pixel with maximum efficiency) was 0.746(85).

\subsection{Rate capability}
\label{sec:rate-capability}

For beam energies corresponding to spots S4 and S7, three measurements at different intensity steps were performed to study the rate capability of the system. The steps were at \num{8e7}, \num{6e8}, and \num{3.2e9}~protons/s. In each case, $10^{10}$ protons were shot at the phantom, thus - after background subtraction - the same number of incoming \glspl{pg} was expected for runs at the same beam energy. The integration was performed using histograms of signal timestamps, such as the one in \cref{fig:TimeStructureOfTheBeam}. For each beam energy, the obtained net spill integrals were consistent within uncertainties, i.e. no measurable dead time of \gls{fee}+\gls{daq} was observed. At the highest energy and intensity step investigated in the whole experiment, the average count rate per channel was \SI{26.4}{kcps}, while the limit of deadtimeless operation given in the ASIC manual is \SI{480}{\kilo\hertz}~\cite{tofpetmanual}.

\subsection{PG depth profiles from experimental data}
\label{sec:pg-depth-profiles}

\begin{figure}[!htb]
\centering
\includegraphics[width = 0.7\textwidth]{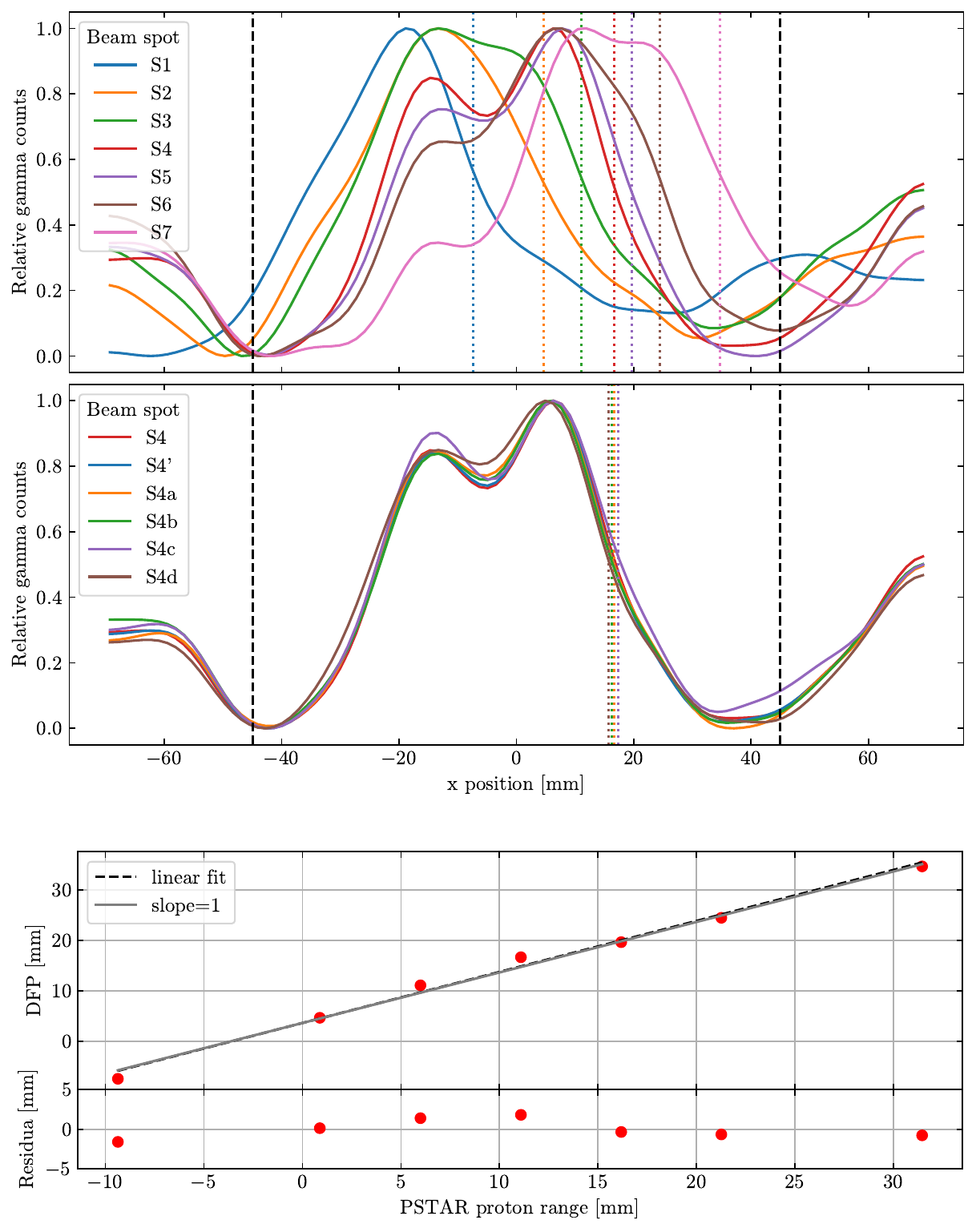}
\caption{\gls{pgh} depth profiles reconstructed from experimental data, smoothed with Gaussian filter. Top panel: profiles for beam spots S1-S7, vertical dotted lines indicate extracted \glspl{dfp}, while the dashed lines mark the phantom edges; middle panel: profiles for beam spots S4, S4', S4a-d (same depth in phantom); bottom panel: distal falloff position vs. proton range, with residuals (distances of the data points from the fitted line). The black dashed line is a linear fit to the data, the grey line is a corresponding line with the same intercept as the fit line, but with the slope equal to 1.}
\label{fig:profiles_exp}
\end{figure}

The number of entries in the hit maps produced from the beam data for \num{e10} protons increased from \num{4.7e5} for the spot S1 to to \num{8.5e5} for spot S7, while the dominating background component, being the intrinsic LYSO:Ce,Ca activity, remained constant. Hence, the background contributions in the spill data were decreasing from \SI{0.50}{\percent} to \SI{0.30}{\percent}.
Based on those hit maps, we reconstructed the \gls{pg} depth profiles for different beam spots (see \cref{fig:IrradiationPlan}). We checked how well the distal falloff positions for all of the profiles correlate with the calculated proton beam range. Based on this correlation, we assessed the accuracy of our reconstruction procedure. The \gls{pg} depth profiles for beam spots S1-S7 are presented in \cref{fig:profiles_exp}~(top panel).

The crucial part of the profile, the distal falloff, is of good quality: its shape is consistent for all beam spots and it contains no major reconstruction artefacts, even though in some profiles changes of steepness can be observed. The rest of the profile, however, presents some features that are unfavourable: although the profiles broaden with increasing depth in the phantom, which is expected, additional wide peaks become visible. This could be due to the Gibbs phenomenon. In addition, it is expected that the rising edge starts at the phantom entry, so that its half-maximum is located close to the phantom border at \SI{-45}{\milli\meter}. We observe the rising edge starting deeper in the phantom; it also shifts to the right with increasing beam energy. Finally, in the regions outside of the phantom, there is an increase in relative gamma counts, even though no such increase is expected from physics. These effects are checked against the simulation in \cref{sec:simResults}. 

The distal falloff positions of these profiles are plotted against the calculated proton ranges in \cref{fig:profiles_exp}~(bottom panel). The reconstruction parameters were optimised according to the scheme described in \cref{sec:performanceMetrics}, yielding the following values:
\begin{itemize}
    \item number of iterations: 23; 
    \item lower threshold of energy deposit in hit maps: \SI{1}{MeV};
    \item \gls{pg} depth profile smoothing with a Gaussian filter, with a kernel standard deviation of 3 pixels;
    \item excluded detector regions: dead pixels and 3 most lateral columns of pixels on both detector sides;
    \item classes of events included: unique and semi-unique.
\end{itemize}
The \gls{rmse} of this set of reconstructions (for $10^{10}$ protons each) is \SI{1.7}{\mm}. The remaining performance metrics are summarised in \cref{tab:performanceMetrics}, along with the corresponding data for simulations, which are presented in detail in \cref{sec:simResults}.

The regions excluded from the analysis were: the dead pixels, the three front (indexed 0-2) and three last (52-54) columns of the detector. The front columns were excluded, because we observed an increased number of counts there due to primary protons from the beam reaching the detector. That was the conclusion drawn based on the spectra of energy deposits in that part of the detector, which extended up to 30~MeV. We excluded the three last columns for symmetry reasons, but also because there was a large fraction of dead pixels there.

The lower hitmap threshold was chosen to be~\SI{1}{MeV}, as it was the most robust, cut out most of the background (see \cref{fig:beam_bg}), but also did not significantly reduce the statistics, which would increase statistical fluctuations.

\begin{table}[!htb]
\centering
\caption{Performance metrics used in the image reconstruction. Full simulation is the simulation with all \glspl{sipm} active, i.e. without the acceptance gaps.}
\label{tab:performanceMetrics}   
\begin{tabularx}{1.0\textwidth}{lXXX}
  \toprule
  ~ & Experiment & Simulation & Full simulation \\
  \midrule
  Correlation coeff. & 0.996012 & 0.998072 & 0.999768 \\ 
  \gls{rmse} [mm] & 1.7 & 1.6 & 0.4 \\ 
  Slope & 1.0102(16) & 0.94351(68) & 0.999981(92) \\
  \bottomrule
\end{tabularx}
\end{table}

\subsection{Analysis of statistical precision}
\label{sec:statistical-precision}

\begin{figure}[!htb]
\centering
\includegraphics[width = 0.99\textwidth]{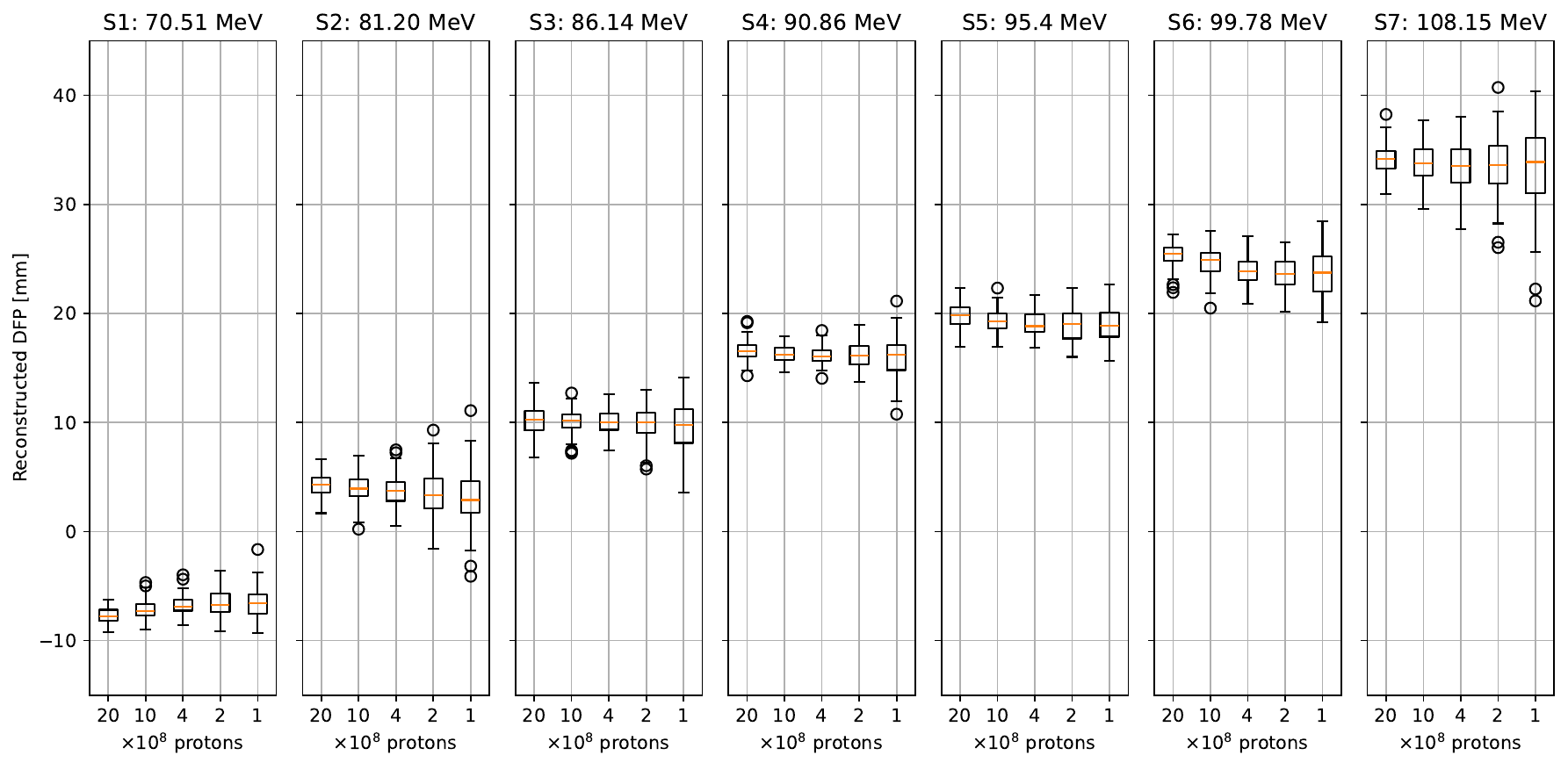}
\caption{Analysis of statistical precision: median and \glspl{iqr} of the reconstructed \glspl{dfp} vs. sample size, for beam spots S1-S7.}
\label{fig:statisticalAnalysis}
\end{figure}

We performed the reconstruction on smaller size data samples, as described in \cref{sec:analysisOfStatisticalPrecision}. The results are presented in \cref{fig:statisticalAnalysis}. There are seven plots for beam spots S1-S7, arranged in ascending beam energy order. The horizontal axis represents the sample size used for image reconstruction, and the vertical axis shows the reconstructed \gls{dfp} in \si{\mm}. For each data point (box) in this plot, 100 reconstructions were performed. The plot type is a standard box plot, i.e. the rectangle (box) is the \gls{iqr} between the quartiles Q1 and Q3. The line denotes the median. The whiskers extend to the furthest data point that falls into a range of $1.5 ~ \cdot$ \gls{iqr}, starting from the box edge. Any data point outside the whiskers range is marked with an empty circle.

For all but the deepest beam spot (S1-S6), the \gls{iqr} is below \SI{3}{\milli\meter}, even for the smallest sample size (\num{e8} protons). The deepest spot (S7) has larger \gls{iqr}, but for the smallest sample size, it is still about \SI{5}{\milli\meter}. The \gls{iqr} increases with decreasing sample size in most cases; however, some exceptions can be observed, e.g. for S3 the IQR for sample size of \num{20e8} is larger than for \num{10e8}, which is counterintuitive. We suspect this effect could be caused by the use of the bootstrapping method. Moreover, we observe that the median value varies slightly for decreasing sample size, it decreases for beam spots S2 and S6, increases for S1, for the remaining beam spots no clear pattern is observed.

\subsection{PG depth profiles from simulation data} \label{sec:simResults}

\begin{figure}[!htb]
\centering
\includegraphics[width = 0.7\textwidth]{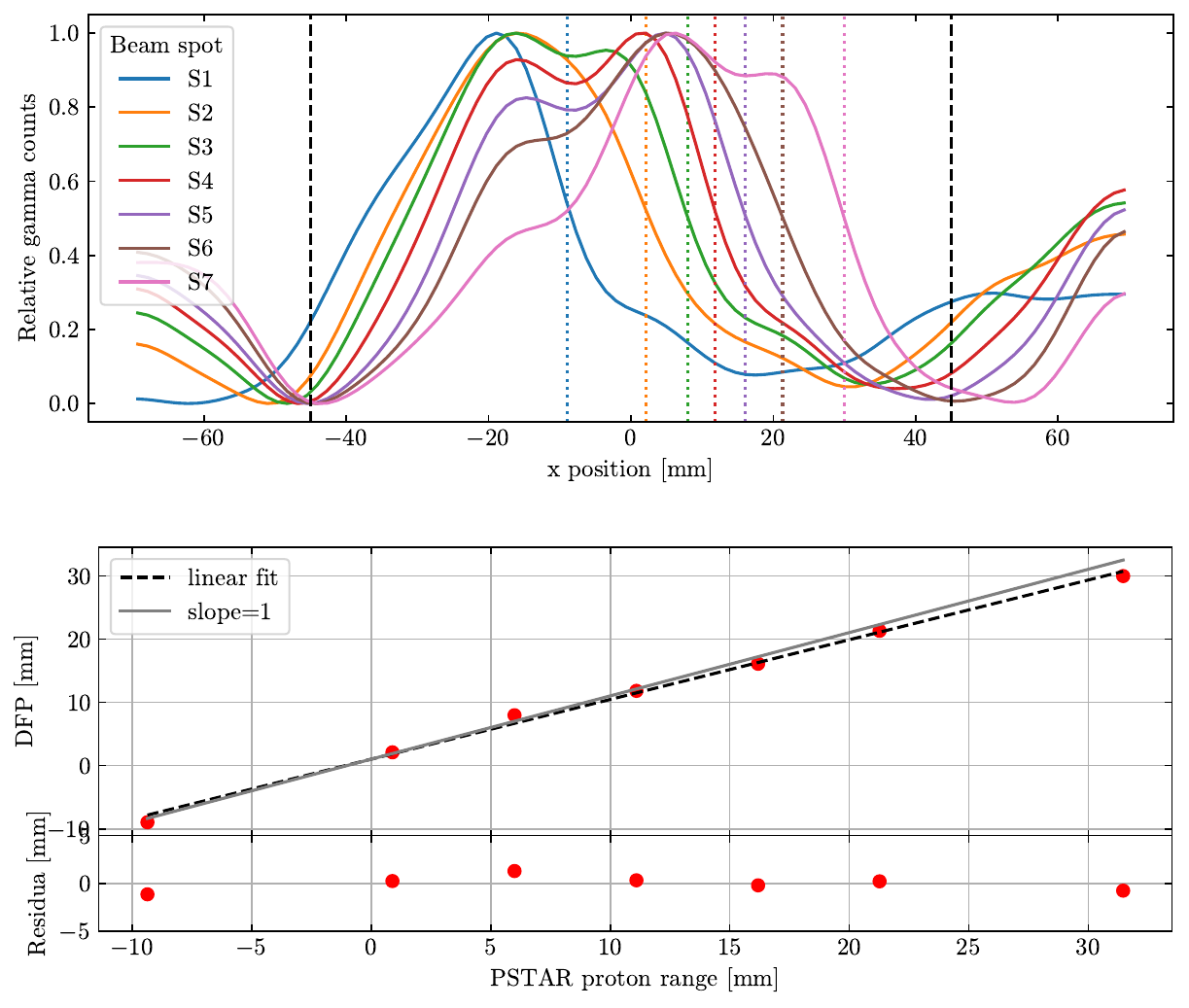}
\caption{\gls{pgh} depth profiles reconstructed from simulated data, smoothed with Gaussian filter. Top panel: profiles for beam spots S1-S7;  bottom panel: distal falloff position vs. proton range, like in~\cref{fig:profiles_exp}.}
\label{fig:profiles_sim}
\end{figure}

To compare the simulation results with the experiment, we reconstructed \gls{pgh} depth profiles from the simulation data. The simulation procedure is detailed in \cref{sec:simulations}. To ensure consistency with experimental conditions, we disabled the \glspl{sipm} that were not operational during the experiment and used the same reconstruction parameters as for the experimental data.
Image reconstruction was performed for beam spots S1-S7, and the results are presented in \cref{fig:profiles_sim}.

The simulated PG depth profiles present features similar to the experimental profiles (see~\cref{sec:pg-depth-profiles}).

\begin{figure}[!htb]
\centering
\includegraphics[width=0.7\textwidth]{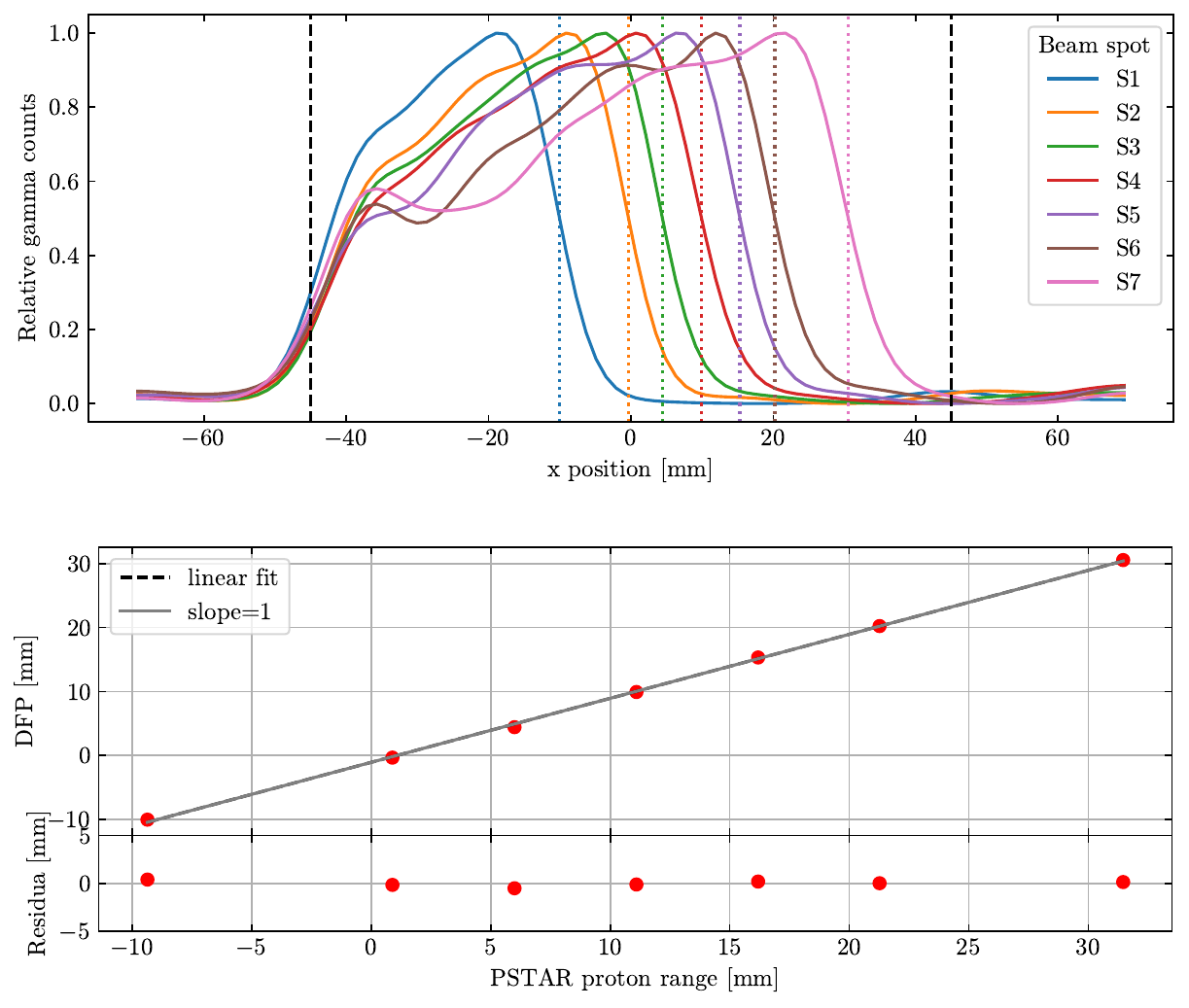}
\caption{\gls{pgh} depth profiles reconstructed from simulated data without acceptance gaps (non-filtered), smoothed with Gaussian filter. Top panel: profiles for beam spots S1-S7;  bottom panel: distal falloff position vs. proton range, like in~\cref{fig:profiles_exp}.}
\label{fig:profiles_sim_nonFiltered}
\end{figure}

To evaluate the impact of acceptance gaps on PG profiles, we performed the reconstruction with all \glspl{sipm} functioning. The corresponding results, shown in \cref{fig:profiles_sim_nonFiltered}, demonstrate that the artefacts described above are now significantly reduced: the reconstructed gamma intensities in the \gls{fov} regions outside of the phantom are low and stable, the rising edge appears earlier and overlaps well with the phantom proximal edge, and the distal falloff shapes are nearly identical for all energies. In this case, the number of iterations was re-optimised and found to be 600, which was higher than for the experimental data. The rest of the reconstruction parameters remained unchanged. 

Performance metrics for these reconstructions are presented in \cref{tab:performanceMetrics}.
\section{Discussion}

\subsection{Experimental and simulated PG depth profiles}

A qualitative comparison of \gls{pgh} depth profiles reconstructed from experimental data and from simulation accounting for most of the experimental effects (\cref{fig:profiles_exp,fig:profiles_sim}, respectively) shows that both sets possess very similar features. A more detailed, quantitative analysis checking the linearity of the \gls{dfp} dependence on the proton range also shows that the resulting RMSE values for both sets are almost the same (see \cref{tab:performanceMetrics}). This gives us confidence that the implemented setup description and the remaining simulation settings in our simulations describe the irradiation experiments accurately and realistically.

The spread of the \glspl{dfp} extracted for all the spots at the depth of S4, including those with modified lateral beam positions and increased number of impinged protons, was not larger than the achieved \gls{dfp} precision. Hence, we conclude that the camera is not sensitive to the variation of the beam lateral position within $\pm\SI{1}{\cm}$.

The investigated setup was not free from construction flaws, i.e. non-working \glspl{sipm}. This reduces the information available for image reconstruction and introduces artefacts in hit maps. To analyse this effect and assess its impact on the quality of beam range assessment, a data set with a fully operational detector was simulated and analysed. When repeating the process of optimisation of image reconstruction for this data set, we noticed that here, the performance metrics favour higher numbers of iterations, even up to 600, while 23 was the optimum for experimental data. For the latter, the acceptance gaps in the detector introduced artefacts in the PG depth profiles, which were amplified for larger numbers of iterations, deteriorating the reconstruction performance. That was not the case for the full-simulation data, free from acceptance gaps. Consequently, the accuracy of range shift determination measured with \gls{rmse} could be improved by a factor of 4 - see \cref{tab:performanceMetrics}. This remains to be confirmed experimentally, with the upgraded readout PCBs.

\subsection{Clinical feasibility}
Rate-wise, the setup is ready to operate at clinical beam intensities of a synchrotron-based therapy centre. Moreover, there is still room to safely increase the rates by a factor of 18 without introducing substantial dead time. This makes the setup operational also at a cyclotron-based centre, in which the instantaneous beam intensities are typically ten times larger than at a synchrotron.

\begin{figure}[!htb]
\centering
\includegraphics[width = 0.7\textwidth]{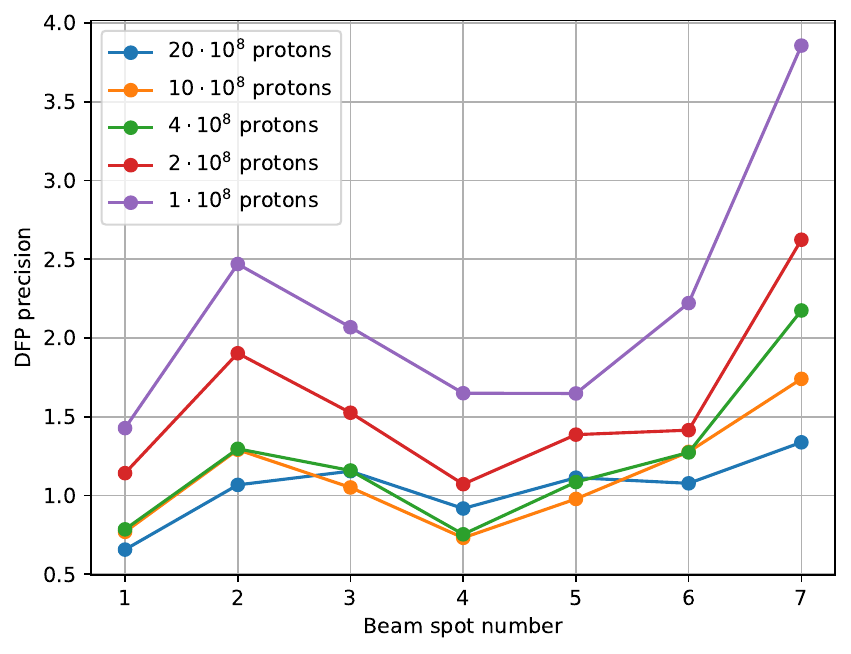}
\caption{Precision of \gls{dfp} determination across the \gls{fov} for different numbers of protons used in phantom irradiation.}
\label{fig:DFPP}
\end{figure}

The studies of statistical precision in \gls{dfp} determination presented in \cref{sec:analysisOfStatisticalPrecision}, translated to the usual metric of standard deviation $(1\sigma)$ are summarized in \cref{fig:DFPP}.
Evidently, the \gls{dfp} precision excels in the region close to the reference spot S4. The precision is not only a function of the statistics used in image reconstruction, but also of the position in the \gls{fov}. The imaging resolution in the most downstream region is about two times worse due to the presence of broader slits in that part of the collimator. Another contribution may come from the neutron background which is expected to be larger at higher beam energies, although the reconstructed profiles do not give a clear supporting evidence.

Our results are the first experimental test of a \gls{cmh} camera for proton therapy monitoring. Other setups, however, have been extensively tested. \Cref{tab:comparisonWorld} compares the performance of the best solutions currently reported in the literature. For the comparison we have selected data from spot S4, as the one that was most extensively investigated, and the number of protons \num{e8}, as it is close to a typical distal spot intensity. As some of the referred publications contain results for multiple beam energies and integrated beam intensities, data for energies and numbers of impinged protons closest to those above have been selected.

\begin{table}[ht]
  \centering
  \caption{Precision $(1\sigma)$ of \gls{dfp}/range estimation by different groups and different \gls{pgi} approaches. All cases but one (see Comments) are 1D imaging.}
  \label{tab:comparisonWorld}
  \small
  \begin{tabularx}{1.0\textwidth}{lllll}
  \toprule
  Approach & $T_\mathrm{p}$  & $N_\mathrm{p}$ & Precision & Comment\\
  & [\si{\MeV}] &  & [\si{\mm}] & \\
  \midrule
  \gls{cm} exp. (this work) & 90.86 & $10^8$ & 1.7 & ref. position in \gls{fov}\\
  \gls{cm} sim. \cite{Hetzel2023} & 85.9--107.9 & $10^8$ & 0.72 & \\
  \gls{cm} sim. \cite{FISTAexp2020} & 122.7 & $10^8$ & 2.1& 2D imaging\\[1ex]
  \gls{mps} exp. \cite{Ku2023} & 99.68 & $10^8$ & 1.5 &\\
  \gls{mps} sim. \cite{pinto_design_2014} & 160 & $10^8$ & 1.30--1.66& \\[1ex]
  \gls{kes} clinical \cite{xie_prompt_2017} & 100--160 & $10^8$ & $\sim2.0$\footnotemark& with spot aggregation\\
  \gls{kes} clinical \cite{Berthold2021} & 160 & \num{1.4e8} & 2.0 & $2^{\mathrm{nd}}$ generation setup\\
  \hline
  \end{tabularx}
\end{table}

\footnotetext[5]{The range of precision 0.7-1.3~mm given by us in Table~2 of \cite{Hetzel2023} was incorrect - it corresponded to the precision of range shifts aggregated over all spots in 9 energy layers, not a statistically-driven precision of a single spot.}

Clearly, the KES setups present much more mature solutions than our camera, are integrated with the therapeutic environment and equipped with support structures offering translational and rotational degrees of freedom enabling precise positioning. They have also been incorporated into the clinical workflow. However, when comparing their achieved \gls{dfp}/range precision achieved per \num{e8}-proton spot, without spot aggregation, our solution performs very similarly, or even somewhat better, despite the flaws in hardware. Removal of the latter, according to the simulation results, is expected to significantly improve the setup precision in range determination, even below the level presented by the currently best performing setup: \gls{mps} of \cite{Ku2023}. The \gls{cm} gamma camera offers also a larger \gls{fov} than KES setups at comparable or smaller material budget.

\subsection{Prospects for 2D imaging}
As shown in \cite{FISTAexp2020} and in our tests with a small-scale prototype~\cite{Hetzel2023}, the \gls{cmh} setup can also be used for 2D imaging. For this purpose, good position resolution is needed in all dimensions. In the case of the tested setup, even though we intended to test this modality with the current setup, the $x$ and $z$ coordinates of the interaction point are reconstructed based on the identification of responding fibers, while the $y$ coordinate needs to be reconstructed from the top/bottom signal ratio. Even though in the R\&D phase the materials of the sensitive part were carefully optimised to achieve the relevant $y$-position resolution, which was verified in measurements with single fibers and small-scale prototypes, the eventually delivered scintillation module presented a much worse performance in this aspect, see \cref{sec:calres}, disabling the 2D imaging option. This, however, will be tested in the future, after the planned reiteration of the construction of the detector sensitive part.

\section{Conclusions}
The presented study demonstrates the feasibility of the \gls{cmh} gamma camera for proton therapy monitoring, providing a novel approach to \gls{pgi} with promising experimental results. We tested the scintillating-fiber-based detector with a coded mask collimator in clinical conditions at \gls{hit}. In the experiment, a \gls{pmma} phantom was irradiated with proton beams of energies ranging from \SI{70,51}{\mega\electronvolt} to \SI{108,15}{\mega\electronvolt} and three intensities ranging from \num{8e7} to \num{3,2e9}~protons/s. The tested setup achieved a precision in \gls{dfp} determination of \SI{1.7}{\milli\meter} with \num{e8} protons at the beam energy of \SI{90.86}{\mega\electronvolt}. Obtained gamma depth profiles were benchmarked with the results of Monte Carlo simulations, showing good agreement. Simulations of the setup without the acceptance gaps (non-filtered) showed that the setup performance can be further improved. In the rate capability studies, we proved that the detector can handle both synchrotron and cyclotron beam intensities without a significant dead time.  Comparison with other existing setups of similar characteristics shows that the \gls{sificc} \gls{cmh} gamma camera is a competitive solution with the potential to be used in clinical practice. Future plans of the collaboration include the elimination of the acceptance holes and exploration of 2D imaging possibilities. 
\ifanonymous
\vspace{5mm}
\else
\section*{Acknowledgements}
The work presented in this paper was supported by the Polish National Science Centre (grants 2017/26/E/ST2/00618 and 2019/33/N/ST2/02780). The exchange of staff and students between Poland and Germany was possible thanks to the support of the Polish National Agency for Academic Exchange (NAWA) as well as the German Academic Exchange Service (DAAD) (project-ID~57562042).
In this context, the project on which this report is based was funded by the German Federal Ministry of Education and Research (BMBF).  Measurements at HIT were financed from the HITRIplus grant (agreement No 101008548) of EU Horizon2020.
Parts of the experimental setup were financed from the program "Excellence Initiative – Research University" at the Jagiellonian University (Research Support Module No. U1U/W17/NO/28). When working on the project, Jorge Roser was partially supported from the DFG Grant COMMA, project number 383681334.
The authors are responsible for the content of this publication.
\fi
\bibliographystyle{paper}
\bibliography{main}

\end{document}